\documentclass[aoas]{imsart}

\RequirePackage{amsthm,amsmath,amsfonts,amssymb}
\RequirePackage[authoryear]{natbib}
\RequirePackage[colorlinks,citecolor=blue,urlcolor=blue]{hyperref}
\RequirePackage{graphicx}
\usepackage{booktabs}
\usepackage{comment}
\usepackage{lscape}
\usepackage{xspace}
\usepackage{threeparttable} 
\usepackage[export]{adjustbox}
\input{cross_ref_multiple}
\newcommand{\tvec}[1]{\text{vec}#1}

\startlocaldefs

\endlocaldefs

\begin{document}

\begin{frontmatter}
\title{Supervised low-rank approximation of high-dimensional multivariate
functional data via tensor decomposition}
\runtitle{Low-rank approximation of high-dimensional multivariate
functional}

\begin{aug}
\author[A]{\fnms{Mohammad Samsul}~\snm{Alam}\ead[label=e1]{malam3@ncsu.edu}\orcid{0000-0002-0602-5861}},
\author[B]{\fnms{Ana-Maria}~\snm{Staicu}\ead[label=e2]{astaicu@ncsu.edu}\orcid{0000-0002-7434-8726}}
\and
\author[A]{\fnms{Pixu}~\snm{Shi}\ead[label=e3]{pixu.shi@duke.edu}}
\address[A]{Department of Biostatistics \& Bioinformatics, Duke University,\printead[presep={\ }]{e1,e3}}
\address[B]{Department of Statistics, North Carolina State University.\printead[presep={\ }]{e2}}
\end{aug}

\begin{abstract}
Motivated by the challenges of analyzing high-dimensional ($p \gg n$) sequencing data from longitudinal microbiome studies, where samples are collected at multiple time points from each subject, we propose supervised functional tensor singular value decomposition (SupFTSVD), a novel dimensionality reduction method that leverages auxiliary information in the dimensionality reduction of high-dimensional functional tensors. Although multivariate functional principal component analysis is a natural choice for dimensionality reduction of multivariate functional data, it becomes computationally burdensome in high-dimensional settings. Low-rank tensor decomposition is a feasible alternative and has gained popularity in recent literature, but existing methods in this realm are often incapable of simultaneously utilizing the temporal structure of the data and subject-level auxiliary information. SupFTSVD overcomes these limitations by generating low-rank representations of high-dimensional functional tensors while incorporating subject-level auxiliary information and accounting for the functional nature of the data. Moreover, SupFTSVD produces low-dimensional representations of subjects, features, and time, as well as subject-specific trajectories, providing valuable insights into the biological significance of variations within the data. In simulation studies, we demonstrate that our method achieves notable improvement in tensor approximation accuracy and loading estimation by utilizing auxiliary information. Finally, we applied SupFTSVD to two longitudinal microbiome studies where biologically meaningful patterns in the data were revealed.
\end{abstract}


\end{frontmatter}
\section{Introduction}
\label{sec:intro}
Recent advancements and widespread availability of next-generation sequencing (NGS) technology have tremendously benefited research in many areas of biomedical research. One such area is microbiome research, which aims to characterize the human microbiome and understand its connection to human health. For example, existing studies have discovered the complicated role played by the gut microbiome in the cases of inflammatory bowel disease (IBS), diabetes, obesity, malnutrition, liver disease, colorectal, cardiovascular, and several neurological disorders \citep{nagalingam2012role,hsiao2013microbiota,goel2014gut,llorente2015gut,mangiola2016gut,al2018gut,cheng2020intestinal,iddrisu2021malnutrition}. As the composition of the host microbiome is dynamic and interacts with the external environment, longitudinal study designs are increasingly being adopted to reflect a growing recognition of the importance of capturing temporal dynamics of the human microbiome \citep{kostic2015dynamics,lloyd2019multi,kodikara2022statistical,ma2023tensor}. \citet{kodikara2022statistical} outlined challenges, namely, inherent complexity, sparsity, over-dispersion, multivariate, temporal variability, and high-dimensionality, in analyzing longitudinal microbiome data (LMD). These challenges are further compounded by the irregular and inconsistent time points at which data is often collected from subjects, adding further complexity to the analysis. 

Functional data analysis (FDA) is one framework used to model longitudinal data with sparse temporal sampling. Several methods under this framework have been devoted to functional principal component analysis (FPCA), the dimensionality reduction of functions, to assist the analysis and interpretation of longitudinal data \citep{yao2005functional}. However, these methods primarily target univariate functional data and often rely on the spectral decomposition of cross-covariance matrices for multivariate functional data, making them computationally intensive in high-dimensional settings such as microbiome sequencing data. Furthermore, most FDA approaches lack a mechanism for reducing the dimensionality of features, such as operational taxonomic units (OTUs) or Amplicon sequence variants (ASVs), which complicates the exploration of the roles played by high-dimensional microbial features.

The high dimensionality in longitudinal multivariate data presents significant challenges to their analysis, especially when features form complex correlation structures such as those found in microbiome data. Dimensionality reduction of features can help us better understand the latent driving factor behind correlated features and reduce the complexity of subsequent analyses. \citet{armstrong2022applications} discussed commonly used dimensionality reduction techniques that rely on the independence assumption in the context of microbiome data analysis, but such an assumption is inapplicable to longitudinal data. To address this issue, \citet{han2023guaranteed} considered functional tensor singular value decomposition (FTSVD), \citet{ma2023tensor} proposed microTensor, and \citet{shi2023time} discussed temporal tensor decomposition (TEMPTED), which do not require independence assumptions. Some of these methods are capable of treating time as a continuous variable and handling missing time points. However, as unsupervised dimensionality reduction methods, they cannot utilize information from auxiliary variables such as subject-level phenotype information.

Supervised dimensionality reduction was introduced by \citet{bair2006prediction} through supervised principal component analysis in the regression framework involving a scalar response variable. This idea was then utilized for multivariate data as supervised singular value decomposition (SupSVD) \citep{li2016supervised} and for univariate functional data as supervised sparse and functional principal component (SupSFPC) analysis \citep{li2016supervised_svd}, respectively. Later, \citet{lock2018supervised} extended SupSVD for tensor data using PARAFAC/CONDECOMP (CP) decomposition and named the approach SupCP. This method can be used to decompose longitudinal microbiome data by formatting the data into a tensor with three modes representing subject, feature, and time respectively. However, it would require all subjects to share the same time points, a strong requirement that rarely holds in practice due to limitations of study designs and missing time points. One example is the early childhood antibiotics and the microbiome study data analyzed in Section \ref{subsec:ecam}. Besides, SupCP treats time as a discrete mode in the tensor rather than as a continuous variable. 

This paper proposes the supervised functional tensor singular value decomposition (SupFTSVD), a dimensionality reduction method for high-dimensional longitudinal data that simultaneously leverages continuous temporal structure and incorporates supervision from auxiliary variables.   
We organize the subsequent sections in the following manner. Section \ref{sec:model} presents the model setting of SupFTSVD, and Section \ref{sec:estimation} describes the details of our EM algorithm to estimate the model components. In Section \ref{sec:prediction}, we introduce how the dimensionality reduction can be transferred from training to testing data and how to predict testing subjects' trajectories based on auxiliary information. Simulation results evaluating the performance of estimation and prediction at different settings are gathered in Section \ref{sec:simulation_study}. In Section \ref{sec:real_data}, we analyze two longitudinal microbiome data sets obtained through Food and Resulting Microbial Metabolites (FARMM) and Early Childhood Antibiotics and the Microbiome (ECAM) studies. Finally, Section \ref{sec:conclusion} presents our conclusion on the proposed method based on the numerical study and data applications.

\section{Modeling framework}
\label{sec:model}
Let $\{(\mathbf{Y}_{ij},t_{ij})_{j=1}^{m_i},\mathbf{x}_i\}; i=1,2,\ldots, n$ be the observed data, where $\mathbf{x}_i\in\mathbb{R}^q$ is a vector of $q$ covariates specific to subject $i$, $\mathbf{Y}_{ij}\in\mathbb{R}^p$ is a vector of $p$ features observed for subject $i$ at time $t_{ij}\in\mathcal{T}$ with $\mathcal{T}$ being a closed interval. 
The $m_i$ time points, $T_i=\{t_{i1},t_{i2},\ldots,t_{im_i}\}$, at which the measurements on subject $i$ are observed, can be either identical or varying across subjects. 
In the context of longitudinal microbiome data, $\mathbf{Y}_{ij}=(Y_{ij}^1,Y_{ij}^2,\ldots,Y_{ij}^p)'$ represents the vector of transformed relative abundance of different bacterial taxa collected from subject $i$ at time $t_{ij}$, and $\mathbf{x}_i$ represents subject-level auxiliary information such as age and body mass index.


We propose to model the observed $\mathbf{Y}_{ij}$'s through a truncated canonical polyadic (CP)  low-rank structure supervised by covariates $\mathbf{x}_i$:

\begin{eqnarray}
Y_{ij}^{b}&=&\sum_{k=1}^{r}\lambda_k\zeta_{ik}\xi_{bk}\psi_k(t_{ij})+\epsilon_{ij}^{b}; b=1,2,\ldots, p,\label{eq:tmodel1}\\
\zeta_{ik}&=&\mathbf{x}_i'\boldsymbol{\gamma}_k+e_{ik}; k\ge 1,\label{eq:tmodel2}
\end{eqnarray}
where for the $k$th component, $(\zeta_{1k},\dots,\zeta_{nk})$ is the subject singular vector, $\boldsymbol{\xi}_k=(\xi_{1k},\dots,\xi_{pk})$ is the feature singular vector, $\psi_k(\cdot)$ is the singular function defined over $\mathcal{T}$ that is assumed to have integrable second derivative, $\lambda_k$ is the singular value, $\boldsymbol{\gamma}_k$ is a $q-$dimensional vector of regression coefficients associated with $\mathbf{x}_i$, and $e_{ik}$ and $\epsilon_{ij}^b$ are mutually independent Gaussian variables with zero mean and variances denoted by $\tau_k$ and $\sigma^2$, respectively. 
To ensure model identifiability, we perform the reparameterization $\boldsymbol{\beta}_k=\lambda_k\boldsymbol{\gamma}_k$ and $U_{ik} = \lambda_ke_{ik}$, and require $||\boldsymbol\xi_k||_2=||\psi_k||_{L_2}=1$, where $||\cdot||_2$ and $||\cdot||_{L_2}$ represent the Euclidean and $L_2$ norms, respectively. 
In further discussion, we refer to \eqref{eq:tmodel1} and \eqref{eq:tmodel2} as the tensor model and subject-loading model, respectively. Plugging the \eqref{eq:tmodel2} into \eqref{eq:tmodel1} leads to the following formulation:

\begin{align}
\label{eq:comb_model}
Y_{ij}^{b}&=\sum_{k=1}^{r}(\mathbf{x}_i^{'}\boldsymbol{\beta}_k)\xi_{bk}\psi_k(t_{ij})+\sum_{k=1}^{r}U_{ik}\xi_{bk}\psi_k(t_{ij})+\epsilon_{ij}^{b},
\end{align}
where $U_{ik}$ has mean $0$ and variance $\sigma_k^2=\lambda_k^2\tau_k$. 
The quantity $r$ is a tuning parameter in the model $(\ref{eq:comb_model})$, which may be pre-specified or data-driven. We call model $(\ref{eq:comb_model})$ the rank$-r$ supervised functional tensor singular value decomposition (SupFTSVD) model. SupFTSVD extends the FTSVD by incorporating ancillary variables in the subject loading $\zeta_{ik}$ to assist the dimensionality reduction. 
The framework of SupFTSVD also encompasses SupCP if $\psi(t_{ij})$'s are functions taking values on a discrete set instead of a continuous interval.

Representation $(\ref{eq:comb_model})$ for a multivariate functional data differs from methods employing the multivariate Karhunan-Lo\`{e}ve (K-L) expansion \citep{happ2018multivariate} in several ways. K-L expansion aims at capturing the temporal trend in the multivariate functions through a set of orthogonal multivariate eigenfunctions, but it does not reduce the dimensionality of features or quantify feature contribution. In large $p$ settings as in microbiome studies, estimation of K-L expansion can be computationally expensive. In contrast, SupFTSVD aims to provide a low-dimensional representation for subject, feature and time modes simultaneously, and uses a set of univariate singular functions to characterize the prominent shared trends across multivariate functional variables. Although it may not perform as well as K-L expansion in capturing the variability in the functions, its dimensionality reduction in subject and feature modes can be valuable for further investigations of the data and biological interpretations.

\section{Estimation of SupFTSVD}
\label{sec:estimation}

\subsection{An EM algorithm}\label{sec:EM}
We propose a maximum likelihood estimation based on the working assumptions that $U_{ik}\sim N(0,\sigma_k^2)$ and $\epsilon_{ij}^{b}\sim N(0,\sigma^2)$. First, we will define a set of notations to assist the description of the estimation procedure. Let $\mathbf{Y}_i=[\mathbf{Y}_{i1},\mathbf{Y}_{i2},\ldots,\mathbf{Y}_{im_i}]$ be a $p\times m_i$ matrix of data observed from subject $i$ at $m_i$ time points. Let $\mathbf{U}_i=[U_{i1},U_{i2},\ldots,U_{ir}]^{'}$ be the vector of random components in the subject loading model. Define matrix $\mathbf{H}_{i}\in\mathbb{R}^{pm_i\times r}$ with $k$th column specified by $\tvec(\boldsymbol{\xi}_k\circ\Psi_{ik})$, where $\Psi_{ik}=[\psi_{k}(t_{i1}),\psi_{k}(t_{i2}),\ldots,\psi_{k}(t_{im_i})]^{'}$, the operator $\circ$ stands for the outer product, and $\text{vec}(\cdot)$ represents the vectorization of a matrix. Let $\mathbf{D}=\text{diag}(\sigma_1^2,\sigma_2^2,\ldots,\sigma_r^2)$ be the covariance matrix of $\mathbf{U}_i$, $V_i=\sigma^2\mathbf{I}_{pm_i}$ be a $pm_i\times pm_i$ diagonal matrix with all diagonal elements equal $\sigma^2$, $\boldsymbol{\beta}_{\mathbf{x}_i}=[\mathbf{x}_i^{'}\boldsymbol{\beta}_1,\mathbf{x}_i^{'}\boldsymbol{\beta}_2,\ldots,\mathbf{x}_i^{'}\boldsymbol{\beta}_r]^{'}$ be a vector specific to subject $i$, and $\mathcal{E}_i=[\boldsymbol{\epsilon}_{i1},\boldsymbol{\epsilon}_{i2},\ldots,\boldsymbol{\epsilon}_{im_i}]$, where $\boldsymbol{\epsilon}_{ij}=[\epsilon_{ij}^{1},\epsilon_{ij}^{2},\ldots,\epsilon_{ij}^{p}]$, be an error matrix. Denote $\mathbf{y}_i=\tvec{(\mathbf{Y}_i)}$ and $\mathbf{e}_i=\tvec{(\mathcal{E}_i)}$ and let $MVN(\boldsymbol{\mu}, \Sigma)$ denote the multivariate normal distribution with mean vector $\boldsymbol{\mu}$ and covariance matrix $\Sigma$. Then we can rewrite the model $(\ref{eq:comb_model})$ in matrix notation as
\begin{align}
\label{eq:mat_trun_model}
\mathbf{y}_i=\mathbf{H}_{i}\boldsymbol{\beta}_{\mathbf{x}_i}+\mathbf{H}_{i}\mathbf{U}_i+\mathbf{e}_i,
\end{align}
where $\mathbf{U}_i\sim MVN(\mathbf{0},\mathbf{D})$ and $\mathbf{e}_i\sim MVN(\mathbf{0},\mathbf{V}_i)$ under the normality assumptions on $U_{i}$ and $\epsilon_{ij}^b$. According to formulation $(\ref{eq:mat_trun_model})$, we have $\mathbf{y}_i\sim MVN(\mathbf{H}_{i}\boldsymbol{\beta}_{\mathbf{x}_i},\mathbf{H}_{i}\mathbf{D}\mathbf{H}_{i}^{'}+\mathbf{V}_i)$ and  $\mathbf{y}_i|\mathbf{U}_i\sim MVN(\mathbf{H}_{i}\boldsymbol{\beta}_{\mathbf{x}_i}$ $+\mathbf{H}_{i}\mathbf{U}_i,\mathbf{V}_i)$. 

Formulation $(\ref{eq:mat_trun_model})$  resembles a classic linear mixed-effect model but their major difference lies in matrix $\mathbf{H}_i$: it consists of latent parameters $\mathbf{\xi}$ and $r$ functional parameters $\{\psi_k(\cdot);k=1,2,\ldots,r\}$. Similar to \citet{lock2018supervised}, who employed an Expectation-Maximization (EM) \citep{green1990use} algorithm to address latent variables, we propose an enhanced EM algorithm that incorporates a smoothness constraint on the functional parameters. This approach leverages the continuity of the data over time and accommodates scenarios where the temporal sampling $T_i$ varies across subjects.

Let $\boldsymbol{\theta}_r=\{\boldsymbol{\beta}_1, \boldsymbol{\beta}_2,\ldots,\boldsymbol{\beta}_r,\boldsymbol{\xi}_{1},\boldsymbol{\xi}_{2},\ldots,\boldsymbol{\xi}_{r},\psi_1(\cdot),\psi_2(\cdot),\ldots,\psi_r(\cdot),\sigma_1^2,\sigma_2^2,\ldots, \sigma_r^2,\sigma^2\}$ be the set of all parameters involved in a rank-r model. Denote $\ell(\boldsymbol{\theta}_r)$ as the log-likelihood of the observed data. Our goal is to estimate $\boldsymbol{\theta}_r$ by maximizing the following objective function using an EM algorithm:

\begin{equation}\label{eq:obj}
    \ell(\boldsymbol{\theta}_r) + \sum_{k=1}^{r}\eta_k||\psi_k||_{\mathcal{H}},
\end{equation}
where $||\cdot||_{\mathcal{H}}$ represents the RKHS norm with Bernoulli polynomial as the reproducing kernel defined in the same way as \cite{han2023guaranteed} to ensure functions $\psi_k$ have square-integrable second derivatives, and $\eta_k$s are tuning parameters controlling the smoothness of functions $\psi_k$. 

\subsubsection{E-step}\label{subsec:E_step} 

The complete log-likelihood of $\mathbf{y}_i$ and $\mathbf{U}_i$ is
\begin{align*}
 \ell_c(\boldsymbol{\theta}_r) 
&\propto \frac{pM}{2}\log(1/\sigma^2)+\frac{n}{2}\sum_{k=1}^{r}\log(1/\sigma_k^2)-\frac{1}{2}\sum_{i=1}^{n}\left(\mathbf{y}_i^{'}\mathbf{V}_i^{-1}\mathbf{y}_i-2\mathbf{y}_i^{'}\mathbf{V}_i^{-1}\mathbf{H}_{i}\boldsymbol{\beta}_{\mathbf{x}_i}\right.\\
&\left.+\boldsymbol{\beta}_{\mathbf{x}_i}\mathbf{H}_{i}^{'}\mathbf{V}_i^{-1}\mathbf{H}_{i}\boldsymbol{\beta}_{\mathbf{x}_i}-2\mathbf{y}_i\mathbf{V}_i^{-1}\mathbf{H}_{i}\mathbf{U}_i+2\boldsymbol{\beta}_{\mathbf{x}_i}\mathbf{H}_{i}^{'}\mathbf{V}_i^{-1}\mathbf{H}_{i}\mathbf{U}_i+\mathbf{U}_i^{'}\Gamma_{i}^{-1}\mathbf{U}_i\right),
\end{align*}
where $M=\sum_{i=1}^{n}m_i$ and $\Gamma_i^{-1}=\mathbf{H}_{i}^{'}\mathbf{V}_i^{-1}\mathbf{H}_{i}+\mathbf{D}^{-1}$.

Starting with an initial value $\boldsymbol{\theta}_r^{(0)}$, let, with some abuse of notations, $\widehat{\boldsymbol{\theta}}_r^{(s)}=\{\widehat{\boldsymbol{\beta}}_1^{(s)}, \widehat{\boldsymbol{\beta}}_2^{(s)},$ $ \ldots, \widehat{\boldsymbol{\beta}}_r^{(s)},$ $ \widehat{\boldsymbol{\xi}}_1^{(s)}, \widehat{\boldsymbol{\xi}}_2^{(s)},\ldots, \widehat{\boldsymbol{\xi}}_r^{(s)},$ $\widehat\psi_1^{(s)}(\cdot), \widehat\psi_2^{(s)}(\cdot),\ldots,\widehat\psi_r^{(s)}(\cdot),\widehat\sigma_1^{2,(s)},\widehat\sigma_2^{2,(s)},$ $\ldots,$ $\widehat\sigma_r^{2,(s)},\widehat\sigma^{2,(s)}\}$ be the current estimate of $\boldsymbol{\theta}_r$. We provide a discussion on the choice of initial value $\boldsymbol{\theta}_r^{(0)}$ in Section \ref{sec:initials}.  At iteration $(s+1)$, the E-step of the proposed estimation computes 
\begin{align}
\label{eq:e_step}
Q(\boldsymbol{\theta}_r;\mathbf{y}_i,\boldsymbol{\theta}_r^{(s)}) = E_{\boldsymbol{\theta}_r^{(s)}}\left[\ell_c(\boldsymbol{\theta}_r)|\mathbf{y}_i\right]+\sum_{k=1}^{r}\eta_k||\psi_k||_{\mathcal{H}},
\end{align}
The conditional expectation in the right-hand side depends on the distribution of $\mathbf{U}_i$ conditional on $\mathbf{y}_i$. We can show that $\mathbf{U}_i|\mathbf{y}_i$ follows a multivariate normal distribution with  mean $E_{\boldsymbol{\theta}_r}(\mathbf{U}_i|\mathbf{y}_i)$ $=\Gamma_i\mathbf{H}_i^{'}\mathbf{V}_i^{-1}\left(\mathbf{y}_i-\mathbf{H}_i\boldsymbol{\beta}_{\mathbf{x}_i}\right)$ and variance $V_{\boldsymbol{\theta}_r}(\mathbf{U}_i|\mathbf{y}_i)=\Gamma_i$. 
Further, with simple algebra, we can show that $E_{\boldsymbol{\theta}^{(s)}}\left(\mathbf{U}_i^{'}\Gamma_{i}^{-1}\mathbf{U}_i|\mathbf{y}_i\right)=\text{tr}(\Gamma_{i}^{-1}\Gamma_i^{(s)})+(\widetilde{\mathbf{U}}_i^{(s)})^{'}\Gamma_{i}^{-1}\widetilde{\mathbf{U}}_i^{(s)}$, where $\widetilde{\mathbf{U}}_i^{(s)}=E_{\boldsymbol{\theta}_r^{(s)}}\left(\mathbf{U}_i|\mathbf{y}_i\right)$ and $\Gamma_i^{(s)}=V_{\boldsymbol{\theta}_r^{(s)}}\left(\mathbf{U}_i|\mathbf{y}_i\right)$. Using these expected values and rearranging the terms, equation $(\ref{eq:e_step})$ has the simplified expression as
\begin{align*}
Q(\boldsymbol{\theta}_r;\mathbf{y}_i,\boldsymbol{\theta}_r^{(s)})&=\frac{pM}{2}\log (1/\sigma^2)+\frac{n}{2}\sum_{k=1}^{r}\log(1/\sigma_k^2) -\frac{1}{2}\sum_{i=1}^{n}\left[
    \left|\left|\mathbf{y}_i-\mathbf{H}_{i}\boldsymbol{\beta}_{\mathbf{x}_i}-\mathbf{H}_{i}\widetilde{\mathbf{U}}_i^{(s)}\right|\right|_{\mathbf{V}_i}^{2}\right.\\
&\left.+\text{tr}\left\{\left(\mathbf{H}_{i}^{'}\mathbf{V}_i^{-1}\mathbf{H}_{i}\right)\Gamma_i^{(s)}\right\}+\text{tr}\left(\mathbf{D}^{-1}\Gamma_i^{(s)}\right)+\left|\left|\widetilde{\mathbf{U}}_i^{(s)}\right|\right|_{\mathbf{D}}^{2}\right]+\sum_{k=1}^{r}\eta_k||\psi_k||_{\mathcal{H}},
\end{align*}
where $||\mathbf{a}||_{\mathbf{B}}^2=\mathbf{a}'\mathbf{B}^{-1}\mathbf{a}$.

\subsubsection{M-step}\label{subsec:m_estimation}
Update $\widehat{\boldsymbol{\theta}}_r^{(s)}$ by
$$\widehat{\boldsymbol{\theta}}^{(s+1)}_r=\underset{\boldsymbol{\theta}_r}{\arg\max}\text{ }Q(\boldsymbol{\theta}_r;\mathbf{y}_i,\boldsymbol{\theta}_r^{(s)})$$
at the $(s+1)$st iteration. Maximizing $Q(\boldsymbol{\theta}_r;\mathbf{y}_i,\boldsymbol{\theta}_r^{(s)})$ with respect to the elements of $\boldsymbol{\theta}_r$ can be achieved through iteratively updating $\boldsymbol{\beta}_k$, $\boldsymbol{\xi}_k$ and $\psi_k(\cdot)$. Specifically, by rearranging terms in $Q(\boldsymbol{\theta}_r;\mathbf{y}_i,\boldsymbol{\theta}_r^{(s)})$, the update of $\boldsymbol{\beta}_k$, $\boldsymbol{\xi}_k$ and $\psi_k(\cdot)$ become the following three optimization problems, respectively:
\begin{align}
\label{eq:mstep_opt}
\hspace{-10pt}\begin{array}{rl}
    (a)&\underset{\boldsymbol{\beta}_k}{\min}\left|\left|\tvec{(\mathbf{R}_{i,k}^{(s)})}-\tvec{(\widehat{\boldsymbol{\xi}}_k^{(s)}\circ\widehat\Psi_{ik}^{(s)})}(\widetilde U_{ik}^{(s)}+\mathbf{x}_i^{'}\boldsymbol{\beta}_k)\right|\right|_2^2,\\
    (b)&\underset{\boldsymbol{\xi}_k}{\min}\left|\left|\tvec{(\widetilde{\mathbf{R}}_{i,k}^{(s)})}-\tvec{(\boldsymbol{\xi}_k\circ\widehat\Psi_{ik}^{(s)})}\sqrt{(\widetilde U_{ik}^{(s)}+\mathbf{x}_i^{'}\widehat{\boldsymbol{\beta}}_k^{(s+1)})^2+\Gamma_{i,kk}^{(s)}}\right|\right|_2^2,\text{ and}\\    
    (c)&\underset{\psi_k\in L^2(\mathcal{T})}{\min}\left|\left|\tvec{(\widetilde{\mathbf{R}}_{i,k}^{(s)})}-\tvec{(\widehat{\boldsymbol{\xi}}_k^{(s+1)}\circ\Psi_{ik})}\sqrt{(\widetilde U_{ik}^{(s)}+\mathbf{x}_i^{'}\widehat{\boldsymbol{\beta}}_k^{(s+1)})^2+\Gamma_{i,kk}^{(s)}}\right|\right|_2^2+\eta_k\left|\left|\phi_k\right|\right|_{\mathcal{H}},    
\end{array}
\end{align}
where $\widehat{\Psi}_{ik}^{(s)}=[\widehat\psi_k^{(s)}(t_{i1}),\widehat\psi_k^{(s)}(t_{i2}),\ldots,\widehat\psi_k^{(s)}(t_{im_i})]^{'}$, $\Gamma_{i,kk}^{(s)}$ is the value of $\Gamma_{i}^{(s)}$ at cell $(k,k)$, $\mathbf{R}_{i,k}^{(s)}\in\mathbb{R}^{p\times m_i}$ obtained using $\mathbf{y}_i$ and $\boldsymbol{\theta}_r^{(s)}$, $\widetilde{\mathbf{R}}_{i,k}^{(s)}$ is a scaled and shifted version of $\mathbf{R}_{i,k}^{(s)}$, and $L^2(\mathcal{T})$ is the space of real-valued functions with squared-integrable second derivatives defined over $\mathcal{T}$. Exact formulas to compute $\mathbf{R}_{i,k}^{(s)}$ and $\widetilde{\mathbf{R}}_{i,k}^{(s)}$ are available in \ref{ssec:mstep_formula}. 

Note that optimizations $(a)$ and $(b)$ of $(\ref{eq:mstep_opt})$ are ordinary least squares problems, and we have closed-form solutions for $\boldsymbol{\beta}_k^{(s+1)}$ and $\boldsymbol{\xi}_k^{(s+1)}$, which we provide in \ref{ssec:mstep_formula} together with the expression of $\sigma_k^{2,(s+1)}$ for all $k$. For $(c)$, a finite-dimensional closed-form solution is available according to the classic Representer Theorem \citep{kimeldorf1971some}. Specifically, the minimzer $\widetilde\psi_k(\cdot)$ has a representation of $\widetilde\psi_k(\cdot)=\sum_{l=1}^{L}\alpha_l\mathcal{K}(\cdot,t_l)$, where $\mathcal{K}$ is the reproducing kernel associated to the Hilbert space $\mathcal{H}\subset L^2(\mathcal{T})$ that satisfies the regulatory conditions: $(i)$ for any $t\in\mathcal{T}$, $\mathcal{K}(\cdot,t)\in\mathcal{H}$ and $(ii)$ for any $\psi\in\mathcal{H}$, $\psi(t)=\left<\psi,\mathcal{K}(\cdot,t)\right>; t\in\mathcal{T}$, and $\{t_1, t_2, \ldots, t_l\}\subset\mathcal{T}$. Details of $\mathcal{K}$ will be provided in Section~\ref{sec:initials}. The problem is subsequently reduced to the estimation of $\boldsymbol{\alpha}=(\alpha_1,\alpha_2,\ldots, \alpha_L)$. Let us define $\mathbf{B}_k^{(s+1)}=\widehat{\boldsymbol{\xi}}^{(s+1)}\otimes \mathbf{A}_k^{(s)}$, where $\mathbf{A}_k^{(s)}$ is an $M\times M$ diagonal matrix defined as $\mathbf{A}_k^{(s)}=\text{diag}(\widetilde{\zeta}_{1k}^{(s)}\mathbf{I}_{m_1},\widetilde{\zeta}_{2k}^{(s)}\mathbf{I}_{m_2},\ldots, \widetilde{\zeta}_{nk}^{(s)}\mathbf{I}_{m_n})$ and $\otimes$ represents the Kronecker product. \citet{han2023guaranteed} showed that $\widehat{\boldsymbol{\alpha}}=\left[\{\mathbf{B}_k^{(s+1)}\}^{'}\mathbf{B}_k^{(s+1)}\mathbf{K}+\eta_k\mathbf{I}_{M}\right]^{-1}\{\mathbf{B}_k^{(s+1)}\}^{'}\mathbf{R}_k^{(s)},$
where $\mathbf{K}$ is an $M\times M$ matrix with elements given by $\mathcal{K}(t,t'); t,t'\in\{t_{11},t_{12},\ldots, t_{1m_i},\ldots,t_{nm_n}\}$ and $\mathbf{R}_{k}^{(s)}\in\mathbb{R}^{pM\times 1}$ obtained by appending $\tvec{(\widetilde{\mathbf{R}}_{i,k}^{(s)})}$. In each update, we also take $\widehat\psi_k^{(s+1)}(\cdot)=\{\sum_{l=1}^{L}\alpha_l\mathcal{K}(\cdot,t_l)\}/\{\sqrt{||\sum_{l=1}^{L}\alpha_l\mathcal{K}(\cdot,t_l)||_{L^2}}\}$ to ensure its unit norm.

We define a stopping rule using the relative change in the objective function defined as $\delta^{(s+1)}=\{Q(\boldsymbol{\theta}_r^{(s+1)};\tvec{(\mathbf{Y}_i)},\boldsymbol{\theta}_r^{(s)})/Q(\boldsymbol{\theta}_r^{(s)};\tvec{(\mathbf{Y}_i)},\boldsymbol{\theta}_r^{(s)})\}-1$. Specifically, we stop the iterative procedure when $\delta^{(s+1)}$ falls below a pre-assigned value, say $\delta_\text{stop}$. 

\subsection{Choice of starting values and reproducing kernel}
\label{sec:initials}
The proposed EM algorithm starts with an initial value, which we denote here by $\boldsymbol{\theta}^{(0)}$. 
Inspired by \citet{han2023guaranteed}, we set the initial value for $\boldsymbol{\xi}_k; k=1,2,\ldots, r$ by performing the singular value decomposition (SVD) of a $p\times M$ matrix constructed from the observed data. Specifically, we perform SVD of $\mathbf{Y}=[\mathbf{Y}_1,\mathbf{Y}_2,\ldots,\mathbf{Y}_n]$ and use the first $r$ resulting left singular vectors as values for $\{\widehat{\boldsymbol{\xi}}_1^{(0)},\widehat{\boldsymbol{\xi}}_2^{(0)},\ldots,\widehat{\boldsymbol{\xi}}_{r}^{(0)}\}$. We then fit a multivariate linear regression (MLR) of $\bar{Y}_{ik}=(\sum_{j=1}^{m_i}\sum_{b=1}^{p}Y_{ij}^{b}\xi_{bk}^{(0)})/m_i; k=1,2,\ldots,r$ on $\mathbf{x}_i$ and use the resulting coefficient vectors as $\widehat{\boldsymbol{\beta}}_1^{(0)}, \widehat{\boldsymbol{\beta}}_2^{(0)},\ldots, \widehat{\boldsymbol{\beta}}_r^{(0)},$ estimated residuals, $\bar{Y}_{ik}-\mathbf{x}_i^{'}\widehat{\boldsymbol{\beta}}_k^{(0)}$; $k=1,2,\ldots,r$, as $\widetilde U_{i2}^{(0)}, \widetilde U_{i2}^{(0)}, \ldots, \widetilde U_{ir}^{(0)}$, and square root of the error variances as $\sigma_1^{(0)}, \sigma_2^{(0)},\ldots, \sigma_r^{(0)}$. For initializing the singular functions, we employ the RKHS regression described in Section \ref{sec:EM} and use $\widehat{\boldsymbol{\beta}}_k^{(0)}$ and $ \widehat{\boldsymbol{\xi}}_k^{(0)}$ to obtain $\widehat\psi_k^{(0)}(\cdot)$ for every $k\in\{1,2,\ldots,r\}$. We then fit a multiple linear regression of $Y_{ij}^{(b)}$ on the initial components constructed by $\{\mathbf{x}_i^{'}\widehat{\boldsymbol{\beta}}_k^{(0)}+\widetilde U_{ik}^{(0)}\}$, $\widehat{\boldsymbol{\xi}}_k^{(0)}$ and $\widehat\psi_k^{(0)}(\cdot)$ and use the residual standard deviation as $\widehat{\sigma}^{(0)}$. 

Throughout the estimation, we treated the kernel $\mathcal{K}$ as given. In our numerical study and data applications, we use the rescaled Bernoulli polynomials as $\mathcal{K}(\cdot,\cdot)$, for which
\begin{align*}
    \mathcal{K}(t,t') = 1+ K_1(t)K_1(t')+ K_2(t)K_2(t')-K_4(|t-t'|); t,t'\in\mathcal{T}
\end{align*}
where $K_1(t)=t-0.5$, $K_2= 0.5[\{K_1(t)\}^2-(1/12)]$, and $K_4(t)= (1/24)[\{K_1(x)\}^4-0.5\{K_1(x)\}^2+(7/240)]$ for any $t\in\mathcal{T}$. At the M-step, we allow $t$ and $t'$ to take values from the set comprised of all distinct observed time points from all the subjects. 

\subsection{Cross-validation for singular function estimation}
We adopt a data-driven cross-validation approach to choose the value for the tuning parameter $\eta_k$ for each component $k$. Denote $g$ as the number of folds in the cross-validation. 
When updating $\psi_k$ at the $(s+1)$st iteration of the M-step, we randomly split the set $\{t_{ij};i=1,2,\ldots,n; j=1,2,\ldots,m_i\}$ into $g$ subsets without replacement. Let $T_k^{(g)}$ be the $g$th subset when updating $\widehat\psi_k^{(s)}(\cdot)$. We construct matrices $\widetilde{\mathbf{R}}_{i,k,g}^{(s)}$ and $\widetilde{\mathbf{R}}_{i,k,g^c}^{(s)}$ by columns of $\widetilde{\mathbf{R}}_{i,k}^{(s)}$ such that $t_{ij}\in T_k^{(g)}$ and $t_{ij}\notin T_k^{(g)}$, respectively. Also, let $\widehat\psi_{k,g^c}^{(s+1)}(\cdot)$ be the updated value of $\widehat\psi_k^{(s)}(\cdot)$ under a given value of $\eta_{k}$ using the data constructed by $\widetilde{\mathbf{R}}_{i,k,g^c}^{(s)}$ across all $i$ and $j$. Let $\rho_{k}^{(g)}(\eta_k)=corr\{\tvec{(\widetilde{\mathbf{R}}_{i,k,g}^{(s)})},\tvec{(\widehat{\widetilde{\mathbf{R}}}_{i,k,g}^{(s+1)})}\}$, where $corr(\mathbf{v}_2,\mathbf{v}_2)$ is the sample correlation between $\mathbf{v}_1$ and $\mathbf{v}_2$, and $\widehat{\widetilde{\mathbf{R}}}_{i,k,g}^{(s+1)}$ has columns defined by $\{\widetilde U_{ik}^{(s)}+\mathbf{x}_i\widehat{\boldsymbol{\beta}}_k^{(s+1)}\}\widehat{\xi}_{bk}^{(s+1)}\widehat\psi_{k,g^c}^{(s+1)}(t_{ij})$ such that $t_{ij}\in T_k^{(g)}$. The optimal value we choose for $\eta_k$ is $\eta_k^{opt} = \arg\min_{\eta_k} g^{-1}\sum_{g=1}^{g}\{\rho_k^{(g)}(\cdot)\}^2$. 
Since the cross-validation is performed at each iteration, to ensure the convergence of the EM algorithm is not affected by the changing $\eta_k$, we only perform this cross-validation for a given number of iterations $s_0$, and use the resulting $\eta_k$ for all iterations$s>s_0$ until convergence.

\subsection{Low-rank approximation of new data}
\label{sec:prediction}
Let $\{(\mathbf{Y}_{j}^{n},t_{j}^{n})_{j=1}^{m},\mathbf{x}^{n}\}$ be the data available for a new subject. The low-rank approximation learned from previous data can be utilized to obtain a low-rank approximation of the new subject's data. Specifically, the low-rank approximation of $\mathbf{Y}^{n}(\cdot)$ can be obtained as $\widehat{\mathbf{Y}}^{n}(\cdot)=(\widehat Y^{n,1}, \widehat Y^{n,2}, \ldots, \widehat Y^{n,p})^\top$, where
\begin{align}
\label{eq:con_predict}
    \widehat{Y}^{n,b}(t)=\sum_{k=1}^{r}\{(\mathbf{x}^{n})^{'}\widehat{\boldsymbol{\beta}}_k+\widehat{U}_k^{n}\}\widehat{\xi}_{bk}\widehat\psi_k(t);\text{ } b=1,2,\ldots,p,
\end{align}
and $\widehat{U}_k^{n}$ is the $k$th element of $\widehat{\mathbf{U}}^{n}=\Gamma^{n}(\mathbf{H}^{n})^{'}(\mathbf{V}^{n})^{-1}(\mathbf{y}^{n}-\mathbf{H}^{n}\boldsymbol{\beta}_{\mathbf{x}^{n}})$. Here, matrices $\Gamma^{n}$, $\mathbf{H}^{n}$, $\mathbf{V}^{n}$, $\mathbf{y}^{n}$ and $\boldsymbol{\beta}_{\mathbf{x}^n}$ have similar forms to the expressions described in Section \ref{sec:estimation} but with the final estimate $\widehat{\boldsymbol{\beta}}_k$, $\widehat{\boldsymbol{\xi}}_{k}$ and $\widehat\psi_k(\cdot)$ from the previous data replacing their counterparts.

Additionally, we can provide a low-dimensional approximation for the unobserved $\mathbf{Y}^{n}(\cdot)$ of new subjects using their observed auxiliary data $\mathbf{x}_n$. For this case, we denote the low-rank approximation by $\widehat{\mathbf{Y}}_a^{n}(\cdot)$, where the subscript $a$ indicates shared auxiliary information. $\widehat{\mathbf{Y}}_a^{n}(\cdot)$ can be obtained as $\mathbf{Y}_a^{n}(\cdot)=(\widehat{Y}_a^{n,1}, \dots, \widehat{Y}_a^{n,p})^\top$, where
\begin{align*}
    \widehat{Y}_a^{n,b}(t)=\sum_{k=1}^{r}\{(\mathbf{x}^{n})^{'}\widehat{\boldsymbol{\beta}}_k\}\widehat{\xi}_{bk}\widehat\psi_k(t);\text{ } b=1,2,\ldots,p.
\end{align*}

\section{Simulation study}
\label{sec:simulation_study}
We performed a Monte Carlo simulation study to assess the performance of our proposed decomposition method. This section describes the simulation settings and corresponding results. 

\subsection{Simulation settings}
Our numerical study used the model $\eqref{eq:tmodel1}$ and $\eqref{eq:tmodel2}$ to generate the data $\{(\mathbf{Y}_{ij},t_{ij})_{j=1}^{m_i},\mathbf{x}_i\}; i=1,2,\ldots, n$ with a finite truncation $r$. We simulated $\{x_{11},x_{21},\ldots,x_{n1}\}$ and $\{x_{12},x_{22},\ldots,x_{n2}\}$ from $U(0,1)$ and $Beta(1,1)$ distributions, respectively. These subject-specific covariates remained fixed over the simulations of a given sample size. We took $\mathbf{x}_{i}=(x_{i1},x_{i2})'$, $\boldsymbol{\gamma}_k=(\gamma_{k1},\gamma_{k2})'$ for $k=1,2,\ldots,r$ and obtained subject loading via $\zeta_{ik}=\mathbf{x}_{i}^\top \boldsymbol{\gamma}_k +e_{ik}$, where $e_{ik}\sim N(0,\tau_k)$.  The feature loading vector, $\boldsymbol{\xi}_{k}$, was generated uniformly from the unit spheres $\mathbb{S}^{p-1}$, with $p=500\gg n$. To generate singular functions, we drew $m_i$ observations from $U(0,1)$ and use them as $t_{ij}$ after sorting in increasing order for every subject $i\in\{1,2,\ldots,n\}$. We then generated the singular functions $\varphi_k(\cdot)$ from $L^2[0,1]$ that is spanned by the basis functions $\psi_1(t)=1$ and $\psi_l(t)=\sqrt{2}\cos((l-1)\pi t); l=2,3,\ldots,10$, which is similar to \citet{han2023guaranteed}. Specifically, $\varphi_k(t)=\sum_{l=1}^{10}\eta_{kl}\psi_{l}(t)$, where $\eta_{kl}\sim U(-1/l,1/l)$. The $b$th element of $\mathbf{Y}_{ij}$ was then generated using $Y_{ij}^{b}=\sum_{k=1}^{r}\lambda_k\zeta_{ik}\xi_{bk}\varphi_k(t_{ij})+\epsilon_{ij}^{b}$, where $\epsilon_{ij}^{b}\sim N(0,\sigma^2)$.

We conducted the simulation study in two setups. The first setup aims at assessing the performance of methods for varying number of subjects $n$, which we take values from $\{30,50,100,200\}$. The second setup aims at assessing the effect of temporal sampling density $m_i$, i.e. the number of time points each subject is observed, which we take values from $\{3,5,8,10\}$. For each setup, we took $\sigma^2\in\{1,4\}$, $\tau_k$ from $\{1,2,5\}$ for the rank-1 model and from $\{(1,1.5),(2,2.25),(3,4)\}$ for the rank-2 model, resulting in 24 combinations. For the first setup, $m_i$ randomly took values from $\{3,4,\ldots,8\}$ with an equal probability. For the second setup, the number of subjects remained fixed at $n=100$. We set $\gamma_1=(1.5,3)'$, $\lambda_1=80$ for the rank-1 model, and $\gamma_1=(1.5,3)'$, $\gamma_2=(2,3.4)'$, $\{\lambda_1=120,\lambda_2=80\}$ for the rank-2 model. We also conducted parallel simulations with $\gamma_k=0$ while other parameter settings are the same, to assess the performance of SupFTSVD when auxiliary variables are not related to the tensor. Each setting was repeated for $100$ Monte Carlo iterations. 

We examine rank-1 and rank-2 models separately since they are not directly comparable. For the rank-2 model, we also look at its components separately when assessing the estimation accuracy of individual loadings. As assessment criteria for estimation, we use mean-squared error (MSE) for $\mathbf{x}_i^{'}\widehat{\boldsymbol{\beta}}_k$, Euclidean norm of the difference between $\widehat{\boldsymbol{\xi}}_k$ and $\boldsymbol{\xi}$, and $L^2$ distance between $\widehat{\psi}_k(\cdot)$ and $\psi_k(\cdot)$, $\int\{\widehat\psi_k(t)-\psi_k(t)\}^2dt$. For FTSVD, $\widehat{\boldsymbol{\beta}}_k$ are obtained by regressing the subject loading onto the auxiliary variables after decomposition is finished.

To assess the accuracy of low-rank approximation for in-sample subjects, we use the coefficient of determination, $R^2$, obtained by regressing the observed data $\mathbf{y}_i$ on its estimated low-rank components  $\widehat{\mathbf{y}}_i^k; k=1,2,\ldots, r$, where $\widehat{\mathbf{y}}_i^k$ is constructed using $\widehat{\boldsymbol{\beta}}_k$, $\widehat{\boldsymbol{\xi}}_k$ and $\widehat{\psi}_k(\cdot)$. Denoted by $\widehat{\mathbf{y}}_i$ is the fitted by value of $\mathbf{y}_i$ from this linear regression, we can obtain the $R^2$ as 
    \begin{equation}\label{eq:R-squared}
    1-\frac{\sum_{i=1}^{n}\|\mathbf{y}_i - \widehat{\mathbf{y}}_i\|_F^2}{\sum_{i=1}^{n}\|\mathbf{y}_i - \bar{y}\mathbf{1}_{pm_i}\|_F^2},
\end{equation}
where $\bar{y}$ is the average of all elements from $\mathbf{y}_i; i=1,2,\ldots,n$, and $\mathbf{1}_{J}$ is a $J\times 1$ vector of ones. 
For out-of-sample prediction, we evaluate using the mean-squared prediction error (MSPE) defined as
\begin{equation}\label{eq:MSPE}
    (1/nb)\sum_{i=1}^{n}\sum_{b=1}^{p}\int\{\widehat{Y}_i^{b}(t)-Y_i^{b}(t)\}^2dt,
\end{equation}
where $\widehat Y_{i}^{b}(\cdot)$ is a predictor of $Y_{i}^{b}(\cdot)$.

\subsection{Simulation results}
\label{subsec:ss_results}
In this section, we present the simulation results comparing the overall quality of dimensionality reduction between SupFTSVD and FTSVD in terms of tensor approximation accuracy both in-sample and out-of-sample. We also present the alignment between mean subject loadings $(\mathbf{x}_i\boldsymbol{\beta}_k)$ and the auxiliary variables. The estimation accuracy of the feature loadings ($\boldsymbol{\xi}_k$) and singular function $(\psi_k)$ can be found in the supplementary materials. 

 \begin{figure}[!htb]
    \centering
    \includegraphics[width=\textwidth]{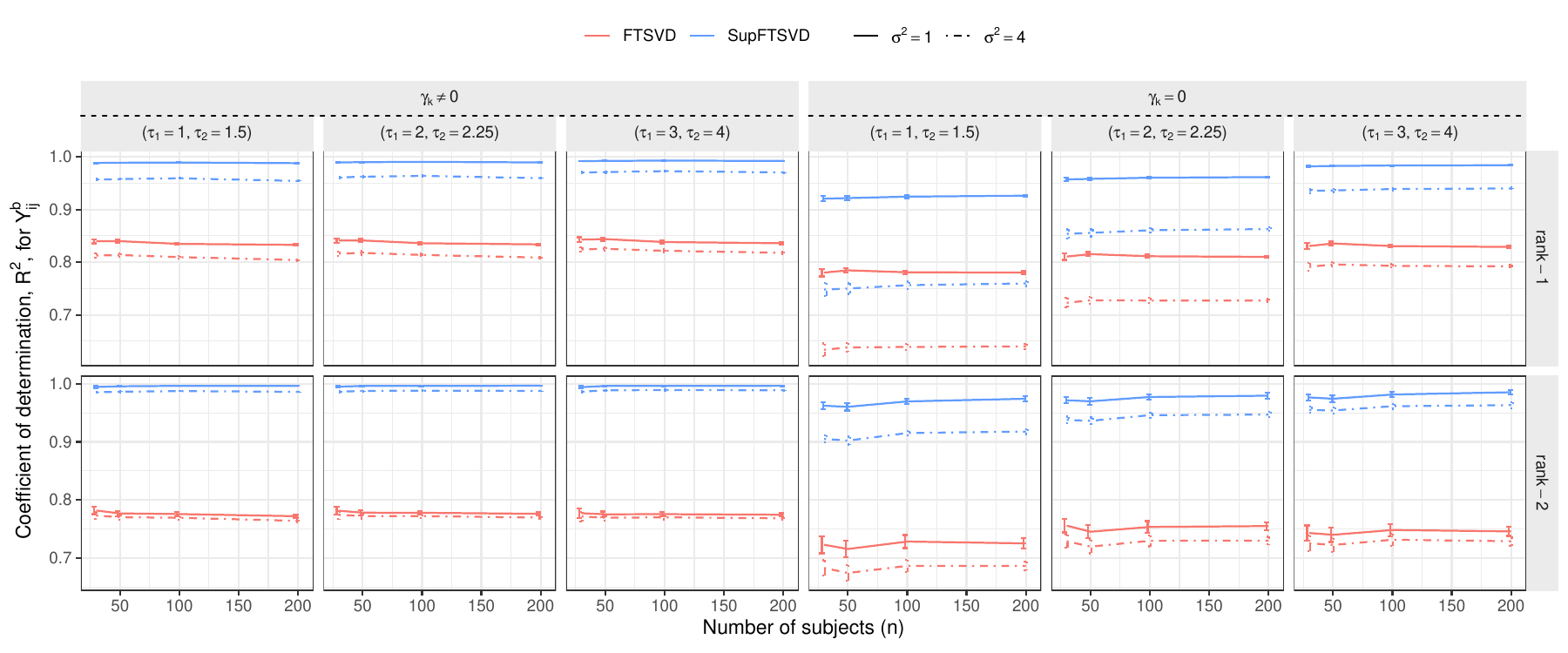}
    \caption{Coefficient of determination $R^2$ \eqref{eq:R-squared} between the observed and estimated data for in-sample tensor approximation accuracy, for different numbers of subjects $n$, with error bars representing two standard deviations on both sides. Results are based on $100$ Monte Carlo samples.}
    \label{fig:r2tensor_n}
\end{figure}

 \begin{figure}[!htb]
    \centering
    \includegraphics[width=\textwidth]{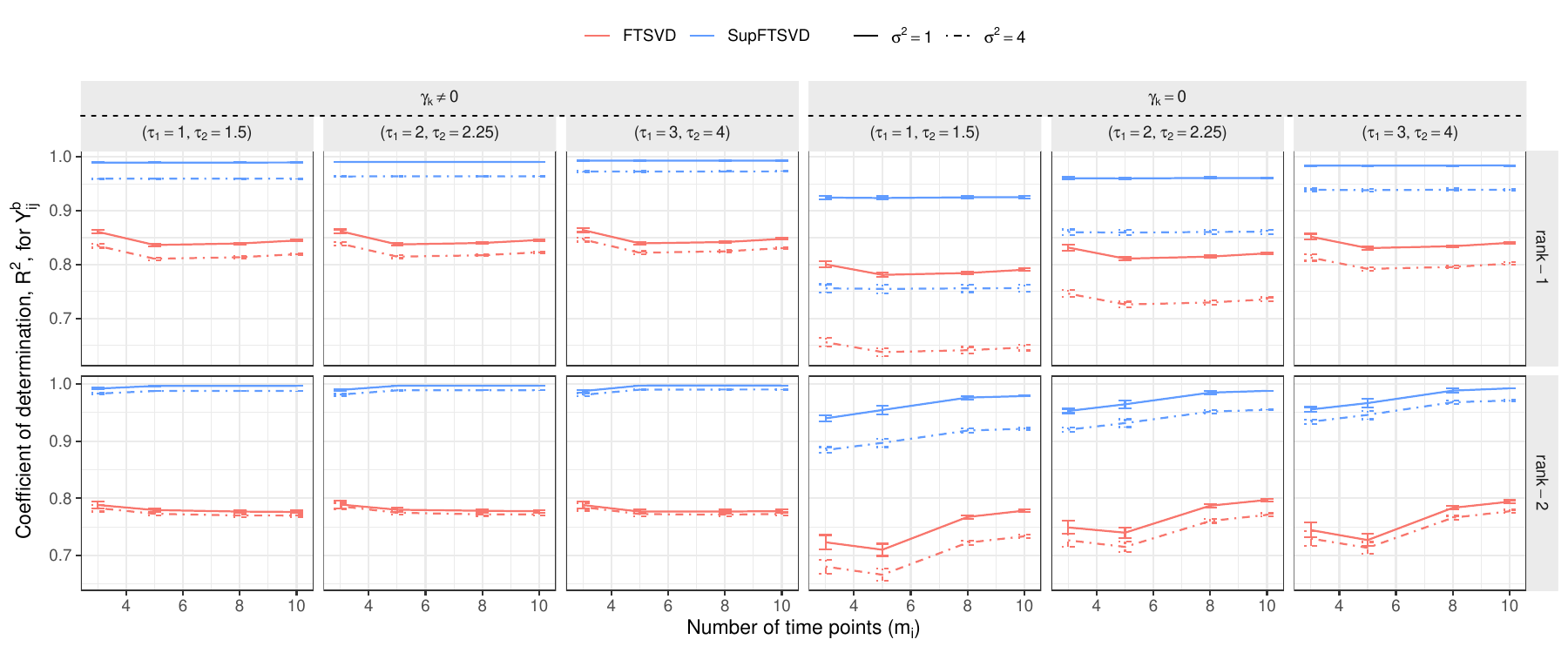}
    \caption{Coefficient of determination $R^2$ \eqref{eq:R-squared} between the observed and estimated data for in-sample tensor approximation accuracy, for different numbers of time points $m_i$, with error bars representing two standard deviations on both sides. Results are based on $100$ Monte Carlo samples.}
    \label{fig:r2tensor_gd}
\end{figure}

Figures \ref{fig:r2tensor_n} and \ref{fig:r2tensor_gd} show the in-sample tensor approximation accuracy in terms of average coefficient of determination $R^2$ \eqref{eq:R-squared}, with error bars representing two standard deviations on each side. SupFTSVD outperforms FTSVD in all settings, indicating that SupFTSVD captures the observed data variation better than FTSVD. For a model of a given rank, we observe little influence from the number of subjects $n$ on $R^2$ for both methods, while the tensor error variance $\sigma^2$ and the subject loading noise level $\tau_k$ both have negative impacts on the performance of methods. Increasing the number of time points $m_i$ improves the performance of both methods when auxiliary variables are irrelevant ($\gamma_k=0$), but has little impact under $\gamma_k\neq0$ because SupFTSVD has already reached an accurate estimation of the singular functions at a small value of $m_i$ whereas FTSVD did not benefit from the increasing $m_i$ because its singular function estimation  was too far from truth (Figure~S4).

\begin{figure}[!htb]
    \centering
    \includegraphics[width=\textwidth]{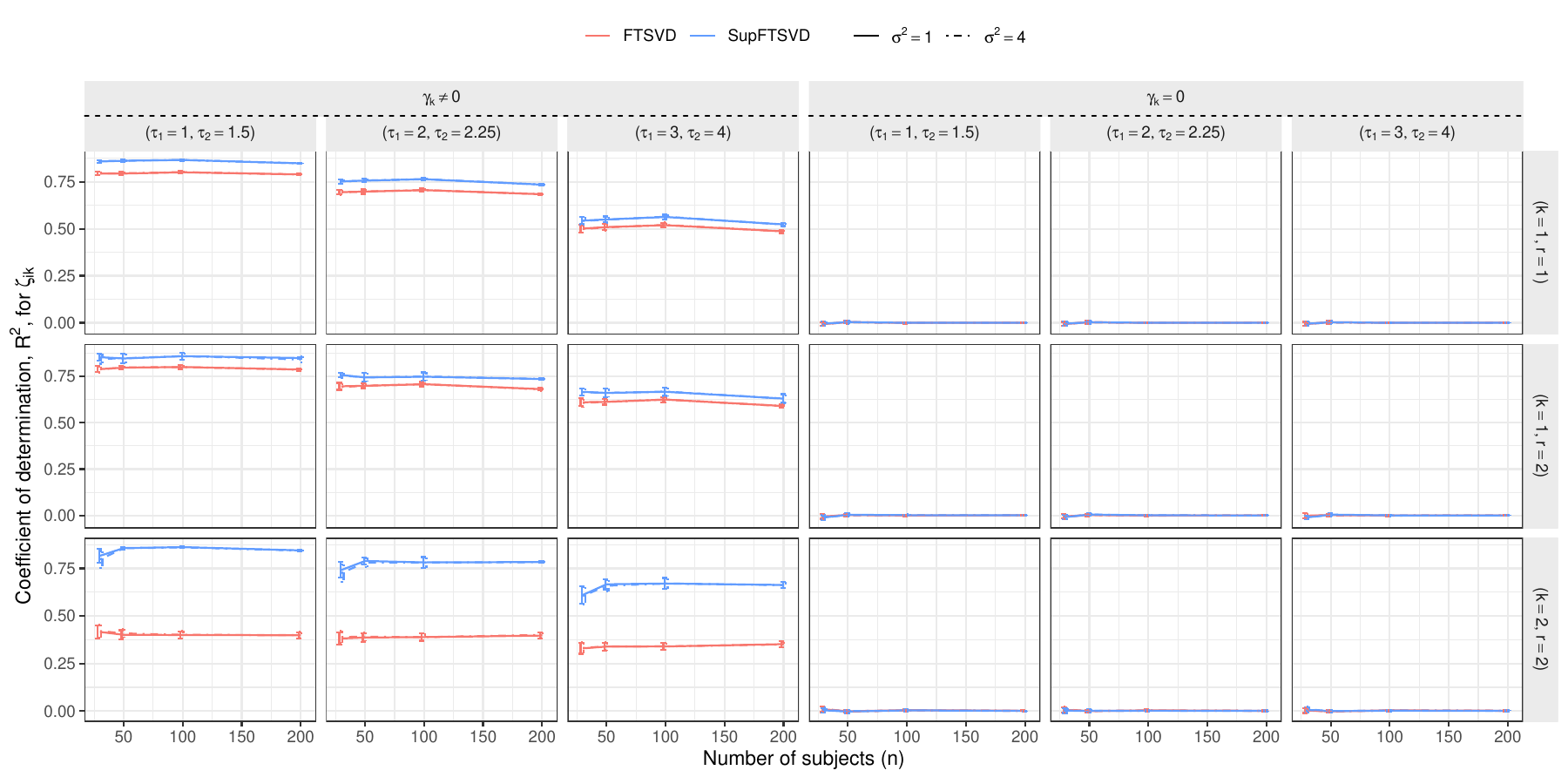}
    \caption{Coefficient of determination, $R^2$, from a linear regression of subject loading $\zeta_{ik}$ on the auxiliary variables $\mathbf{x}_i$ across different numbers of subjects $n$, along with an error bars constructed by 2 times the standard deviation on both sides. Results are based on $100$ Monte Carlo samples.}
    \label{fig:r2loading_n}
\end{figure}

\begin{figure}[!htb]
    \centering
    \includegraphics[width=\textwidth]{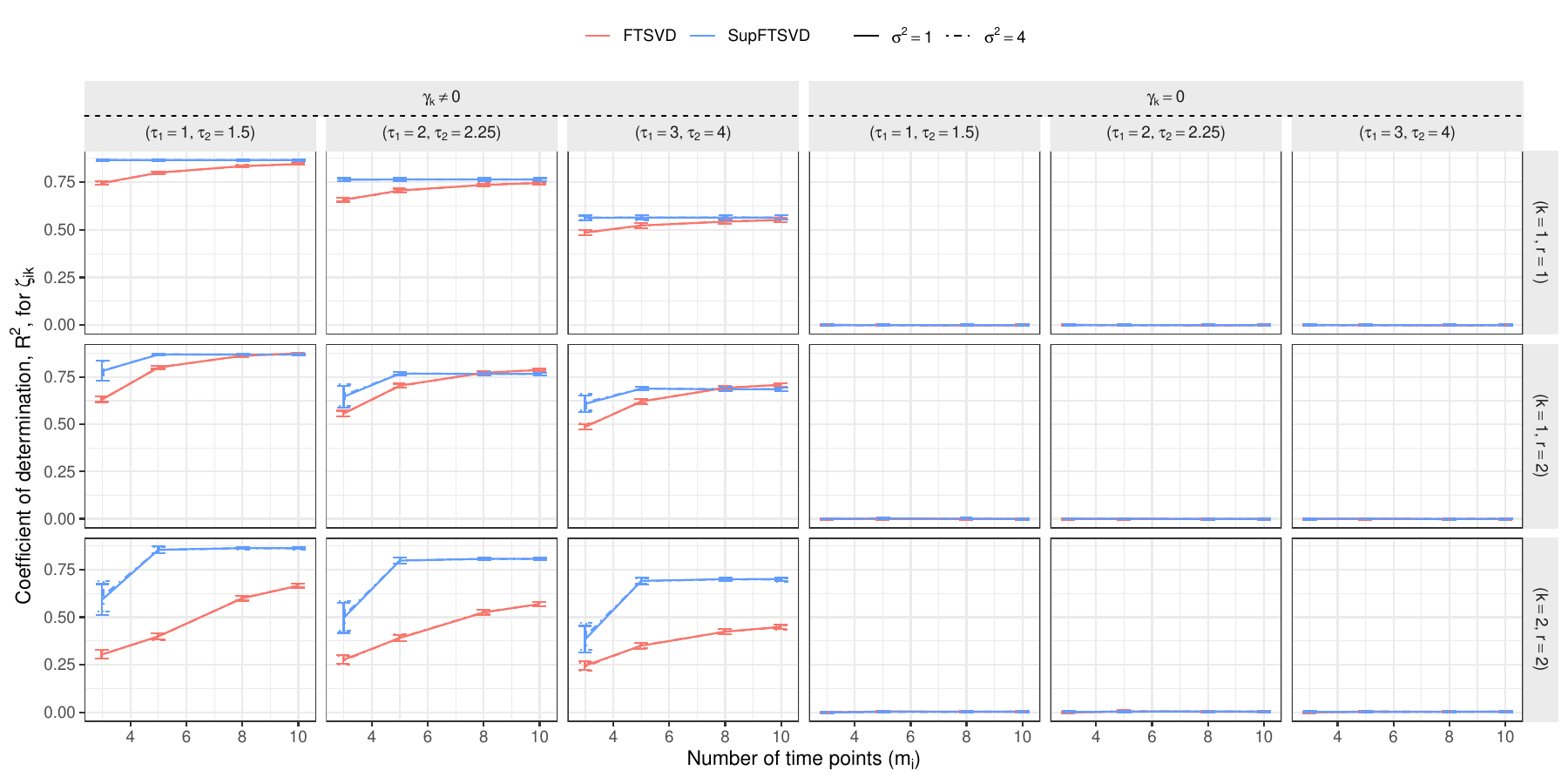}
    \caption{Coefficient of determination, $R^2$, from a linear regression of subject loading $\zeta_{ik}$ on the auxiliary variables $\mathbf{x}_i$ across different numbers of time points $m_i$, along with an error bars constructed by 2 times the standard deviation on both sides. Results are based on $100$ Monte Carlo samples.}
    \label{fig:r2loading_gd}
\end{figure}

Based on the model settings of FTSVD and SupFTSVD, we expect subject loadings from SupFTSVD to be more aligned with the subject-level covariates. We evaluate this using the $R^2$ from regressing the subject loadings on the auxiliary variables, which are summarized in Figure \ref{fig:r2loading_n} and \ref{fig:r2loading_gd}. We can see that when auxiliary variables are irrelevant ($\gamma_k=0$), the $R^2$ of both methods are close to zero under every parameter setting, indicating SupFTSVD is not overfitting the subject loading with auxiliary variables. In contrast, when the auxiliary variables are relevant to the low-rank structure of the tensor ($\gamma_k\neq0$), SupFTSVD produces subject loadings that contain more agreement with auxiliary variables. The advantage of SupFTSVD is more distinctive in the second component of the rank-2 model, indicating that SupFTSVD prioritizes the extraction of variation related to auxiliary variables, even after the first component already captures majority of this variation, all while maintaining superiority in the overall tensor approximation accuracy. We also observe that the alignment of subject loadings with the auxiliary variable is unaffected by the tensor noise $\sigma^2$, but negatively impacted by higher subject-loading noise $\tau_k$, the main contributor to the correlation between auxiliary variables and the subject loadings.

 \begin{figure}[!htb]
    \centering
    \includegraphics[width=\textwidth]{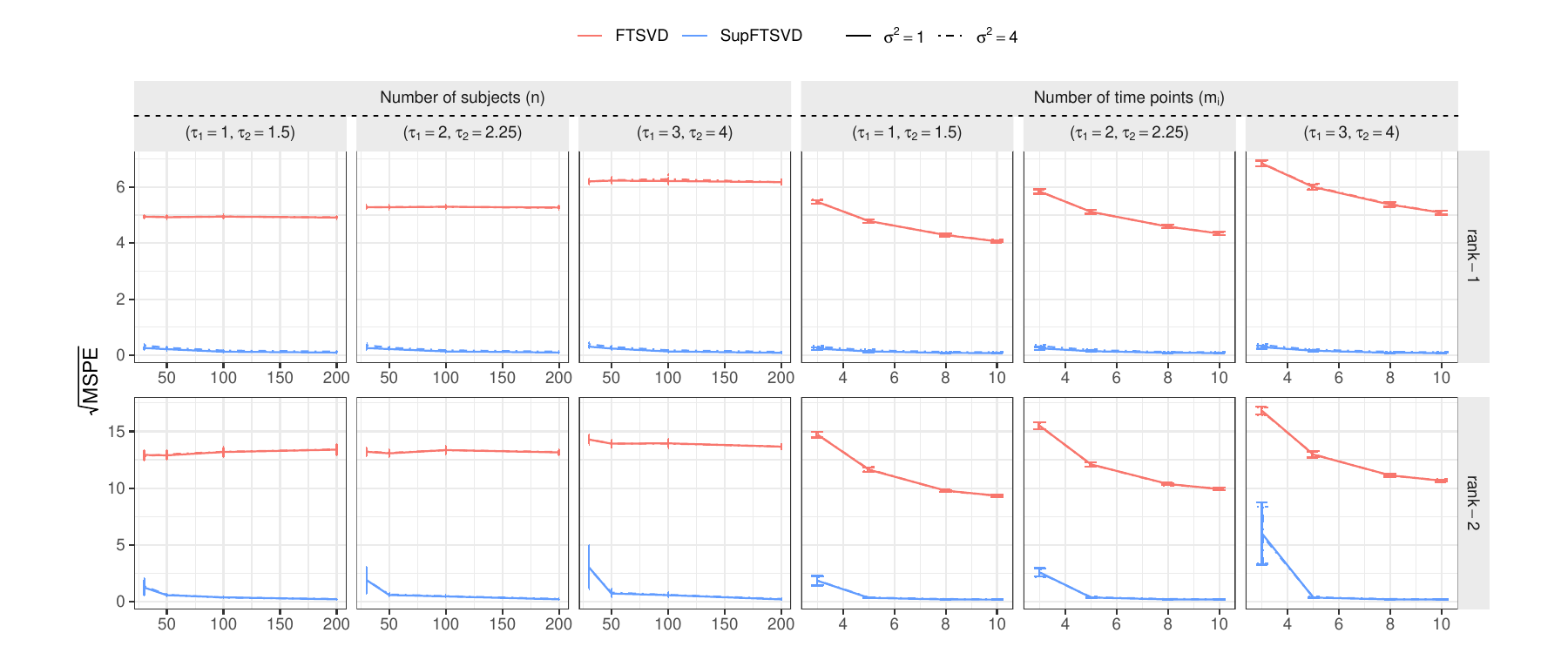}
    \caption{Square root of the mean squared prediction error (MSPE) $\eqref{eq:MSPE}$ of low-rank tensor approximation across different numbers of subjects $n$ and of time points $m_i$ in the case of in-sample and out-of-sample subjects.}
    \label{fig:mspe_n_m}
\end{figure}

In Figure \ref{fig:mspe_n_m}, we present the MSPE \eqref{eq:MSPE} for out-of-sample tensor prediction by SupFTSVD and FTSVD. We obtain these values by performing the dimensionality reduction on the training data of a given number of subject first and then predict for test data consisting of $100$ subjects. To be comparable with FTSVD, we consider both longitudinal data and auxiliary variables are available for the test data, and use Equation \eqref{eq:con_predict} for SupFTSVD. For FTSVD, a similar approach is applied where we inherit the feature loading and singular function from the training data and obtain the subject loadings of the testing data by a one-step linear regression. Here, SupFTSVD achieves substantially better performance compared to FTSVD for every case we considered. We observe an increase in performance when the number of time points $m_i$ increases, likely due to improving estimation of the singular functions. A higher number of subjects $n$ only improves the performance of SupFTSVD when $n$ is small.

\section{Application to longitudinal microbiome data}
\label{sec:real_data}
We use our method to analyze two longitudinal microbiome data sets, the food and resulting microbial metabolites (FARMM) and early childhood antibiotics and microbial (ECAM), published by \citet{tanes2021role} and  \citet{bokulich2016antibiotics}, respectively. Both have subject-level covariates and the counts of operational taxonomic units (OTUs) summarized through bioinformatic analysis of next-generation sequencing data. Previous studies involving low-rank decomposition of these data did not use the available covariates to supervise the decomposition. However, they investigated the resulting components in respect of the covariates. For example, \citet{han2023guaranteed} discussed the temporal dynamics of microbial communities for different delivery modes after analyzing the ECAM data via the FTSVD, and \citet{ma2023tensor} regressed the estimated subject loadings obtained from the analysis of the FARMM data via the microTensor on available subject-level covariates. We aim to represent these longitudinally collected high-dimensional sequence data through a few components capable of showing prominent cross-subjects and temporal variations separately, and connect bacteria to these variations.

16S sequencing data of microbiome samples often possess the following properties: (i) sequencing depth varies across subjects in next-generation sequencing data, and (ii) OTU counts from a microbial study tend to have a right-skewed distribution. Therefore, it is customary to standardize the observed data before any downstream analysis \citep{han2023guaranteed,shi2023time}. We apply a modified version of the centered-log-ratio (CLR) transformation \citep{shi2022high} to the observed counts, $Y_{ij}^{b}$, as $\log[(Y_{ij}^{b}+0.5)/\{\prod_{b=1}^{p}(Y_{ij}^{b}+0.5)\}^{1/p}].$ We also filter low-abundance OTUs before applying our SupFTSVD method to obtain low-rank components. We discuss the OTU filtering criterion later for each of the data applications.

\subsection{Food and resulting microbial metabolites study}
\label{subsec:farmm}
There are $30$ subjects, $10$ for each diet group (EEN, Omnivore, and Vegan), in the FARMM data. The diet EEN stands for exclusive enteral nutrition and represents a liquid diet without fiber that is often used to treat inflammatory bowel disease (IBD). The study collected stool microbial samples daily from every subject over 15 days, with all subjects receiving antibiotics and polythelyne glycol on days $6$, $7$, and $8$ to induce a temporary reduction of bacterial load. Therefore, the entire duration consists of three phases: pre-antibiotic, antibiotic, and post-antibiotic. As SupFTSVD uses time-independent covariates only, we do not take the antibiotic status as a covariate and instead track the changes of microbiome over the course of antibiotic usage.  
\begin{figure}[!htb]
    \centering
    \includegraphics[height=0.6\textheight,width=\textwidth]{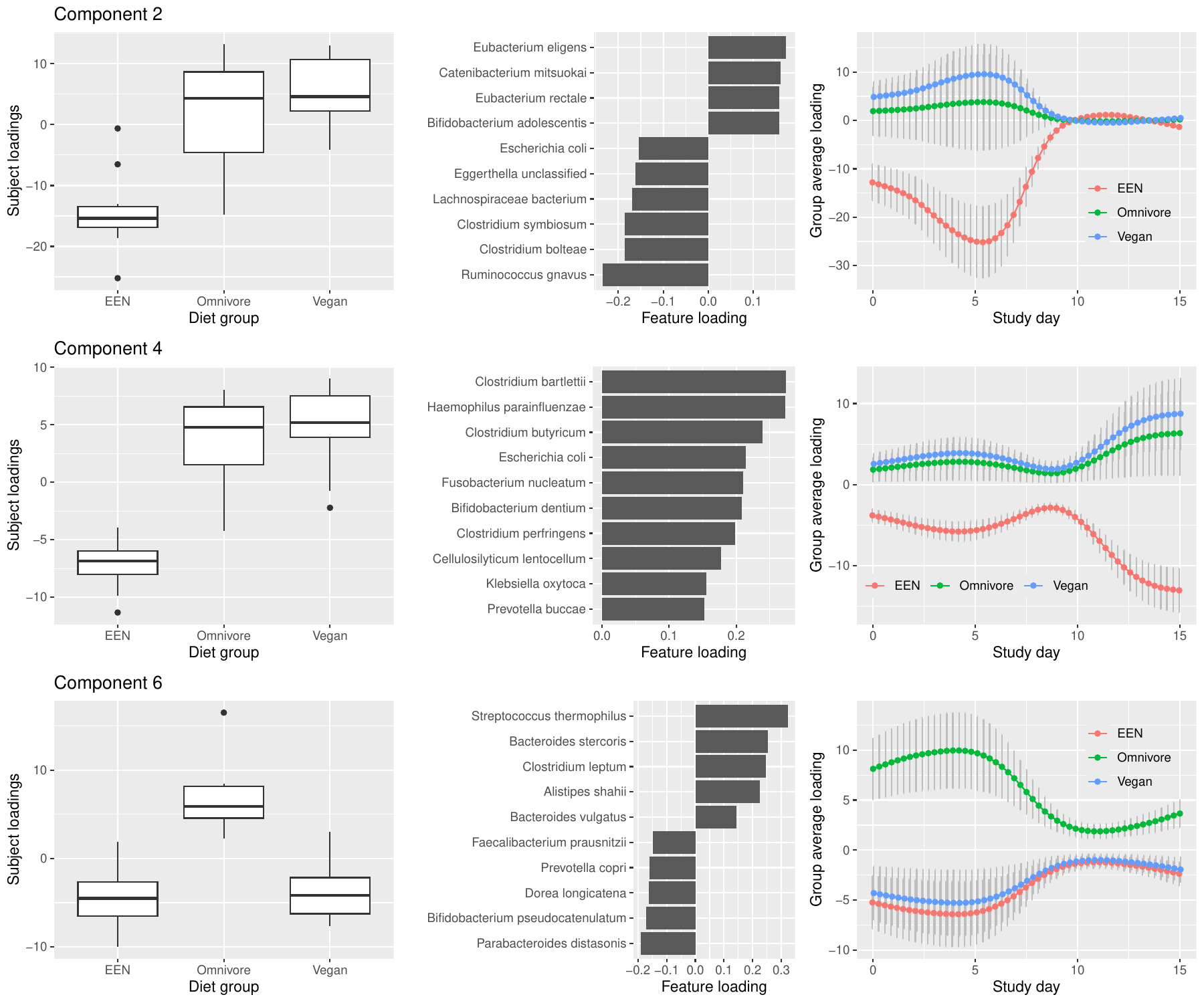}
    \caption{Distribution of subject loadings (left), the mean trend over the study period (right) for three diet groups in the FARMM data, and features with largest loaidngs (middle) for three rank-1 components obtained by fitting a rank-6 model.}
    \label{fig:farm_res}
\end{figure}

The processed FARMM data was obtained from  \citet{ma2023tensor}, where OTUs are filtered out if their relative abundance is lower than $10^{-5}$ relative abundance in at least five samples, which resulted in a feature dimension equal to $343$. In total, we analyze $417$ samples from $30$ subjects. We use subject-specific Age and BMI alongside the diet group to supervise the decomposition in each component. 

Figure \ref{fig:farm_res} shows the predicted subject loadings (in the left column), top ten features according to the extent of estimated loadings (in the middle), and group-wise average singular functions (in the right column) associated with three estimated components (second, fourth, and sixth) from a fitted rank-6 model by the SupFTSVD. To obtain the group-wise average trajectories, we first multiply the estimated singular functions with predicted subject loadings and then take the average of the resulting subject-specific trajectories for each groups. Here we only present the components exhibiting notable associations with dietary groups. The other components may reflect variations associated with variables not collected in this study. Specifically, the boxplots corresponding to components 2 and 4 indicate that subjects in the EEN diet group differ from those in the others. For component 6, we observe a difference between subjects in the Omnivore group and those in the two other diet groups. Point-wise error bars plotted with the average singular functions in the right column further illustrate these group differences. Note that the singular functions discover behaviors across the study period. For instance, during the post-antibiotic period, subjects behave similarly for component 2, whereas discordantly for components 4 and 6. The bar plots presented in the middle column summarize the feature loadings of the dominating genus and species driving the temporal dynamic behind each of these components.

The bacteria our method identifies are consistent with those reported in the literature for their connection with diet and antibiotics. For example, \citet{tanes2021role} reported \emph{Ruminococcus gnavus} as a distinguishing one for the EEN diet, and \citet{gevers2014treatment} showed changes of \emph{Bacteroidales} and \emph{Clostridiales} among the subjects who received antibiotics. 
\citet{perler2023role} is an excellent resource to understand the connection of most bacteria we report here with diet and host health. 

\subsection{Early childhood antibiotics and the microbiome study}
\label{subsec:ecam}
The early childhood antibiotics and microbial (ECAM) study observed 18 to 45-year-old healthy pregnant women longitudinally for three years starting from December 2011. \citet{bokulich2016antibiotics} published the data, and processed data are available to download at \url{https://codeocean.com/capsule/6494482/tree/v1}. We analyze the latter by the proposed SupFTSVD with diet (breastfeeding and formula) and delivery mode (vaginal birth and c-section) as covariates for supervising the low-rank components. 

\begin{figure}[!htb]
    \centering
    \includegraphics[height=0.6\textheight,width=\textwidth]{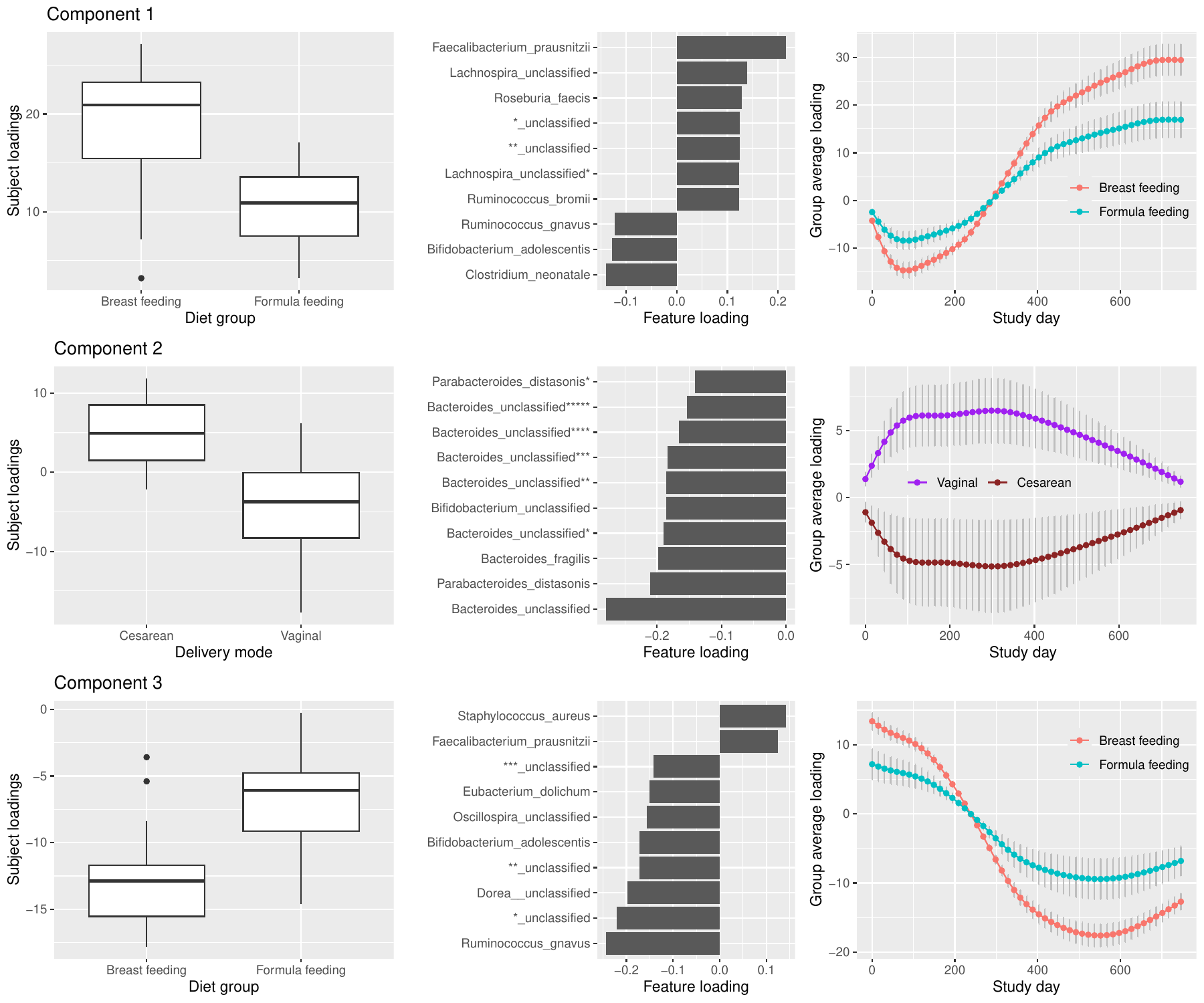}
    \caption{Distribution of subject loadings (left), group-wise mean trends over the study period (right), and features with the largest loadings (middle) for three rank-1 components obtained by fitting a rank-6 model.}
    \label{fig:ecam_res}
\end{figure}
The study collected fecal samples monthly and bi-monthly in the first and second years, respectively. We focus our analysis on the 42 babies with more than one fecal sample during their first two years of life. More specifically, out of $18$ infants delivered by c-section, the numbers of breast and formula-feeding infants are $10$ and $8$, respectively, while these numbers are $20$ and $4$ for the $24$ vaginally delivered infants. Before analyzing, we exclude the OTUs that appear in less than $5\%$ of the samples and fit the SupFTSVD on the remaining $796$ OTUs.

We summarize the results for ECAM data analysis in Figure \ref{fig:ecam_res} similar to that of FARMM data. As we exploit two dichotomous covariates for supervising a rank-6 model, we investigate if any of the estimated components achieves a separation between categories of each of these covariates separately. Specifically, we find that diet groups are connected with components $1$ and $3$, whereas the feeding groups are connected with component $2$. In contrast, the unsupervised decomposition of this data is capable of distinguishing vaginal and cesarean babies but not the diet groups \citep{han2023guaranteed,shi2023time}. Our investigation reveals that \emph{Bacteroides}, \emph{Bifidobacterium}, and \emph{Parabacteroids} dominate the difference between the two delivery modes, which is consistent with the findings of \citet{rutayisire2016mode} and \citet{reyman2019impact}. The results in the first and third rows of Figure \ref{fig:ecam_res} are also supported by previous studies, where relative abundances of \emph{Bifidobacterium}, \emph{Lachnospira}, \emph{Staphylococcus}, and \emph{Ruminococcus} are associated with different feeding modes \citep{guaraldi2012effect,fehr2020breastmilk,martinez2024influence}.

\section{Discussion}
\label{sec:conclusion}
In this paper, we introduced SupFTSVD, a supervised dimensionality reduction method tailored for high-dimensional longitudinal data commonly observed in longitudinal omic studies. SupFTSVD models the observed data as a high-dimensional functional tensor and provides a low-dimensional representation of the data through a CP-type low-rank decomposition. This approach simultaneously characterizes the functional nature of the data and leverages subject-level auxiliary variables to enhance its interpretability. We devised an efficient EM algorithm to estimate the components of the low-rank decomposition, avoiding the computational complexities of K-L expansion-based methods. SupFTSVD also offers the ability to transfer the learned dimensionality reduction from the training to the testing data, promoting research reproducibility. 

Recently, \citet{guan2023smooth} proposed the smoothed probabilistic PARAFAC model with covariates (SPACO) that aims at similar goals. SPACO extends SupCP to have smoothness in the time domain and sparsity on the influence of auxiliary variables, achieved through a difference-based roughness penalty on functions and an $L_1$ penalty on features. The proposed SupFTSVD differs from SPACO by representing the functional parameters using a reproducing kernel Hilbert space (RKHS) similar to \citet{han2023guaranteed}. Both SupFTSVD and SPACO use their respective Expectation-Maximization (EM) algorithms to estimate model components. However, the RKHS representation used by SupFTSVD allows for the updating of temporal components using analytical formulas at the M-step, whereas SPACO requires iterations nested inside the EM algorithm to estimate its temporal components at each M-step, making it more computationally cumbersome than SupFTSVD.  In addition, SupFTSVD has the advantage of transferring dimensionality reduction from training to testing data and predicting new subjects' trajectories based solely on auxiliary variables.

Through simulation studies, we demonstrated the superiority of SupFTSVD over its unsupervised counterpart, showing consistent improvements in both approximation accuracy and the interpretability of the low-rank representation. In the analyses of two real-world longitudinal microbiome studies, SupFTSVD delivers low-rank decompositions of the observed data with components linked to the provided auxiliary variables and identifies bacteria associated with these variables, providing valuable biological insights. We are confident that SupFTSVD will prove to be an important tool for researchers working with high-dimensional longitudinal data.

\bibliographystyle{imsart-nameyear} 
\bibliography{PHD_FUNDATA_2NDGEN}

\setcounter{table}{0}
\renewcommand{\thetable}{A\arabic{table}}
\setcounter{figure}{0}
\renewcommand{\thefigure}{A\arabic{figure}}
\setcounter{section}{0}
\renewcommand{\thesection}{Appendix A.\arabic{section}}
\section{Theoretical derivation}
This section contains detailed derivation of complete data likelihood, conditional mean, and variance besides additional simulation results. We organize this material in four Sections ordered as follows. Section \ref{sec:likelihood} shows the derivation of complete data likelihood, Section \ref{sec:con_mean_var} shows how we obtain the expression for conditional mean and variance of subject loading vector, and Section \ref{ssec:mstep_formula} shows the closed-form solutions for updating different parameters.


\subsection{Complete data likelihood}
\label{sec:likelihood}
In matrix notation, the rank-r model becomes
\begin{align}
\label{eq:mat_trun_model}
\mathbf{y}_i=\mathbf{H}_{i}\boldsymbol{\beta}_{\mathbf{x}_i}+\mathbf{H}_{i}\mathbf{U}_i+\mathbf{e}_i,
\end{align}
where subject-specific random effects $\mathbf{U}_i$ and measurement errors $\mathcal{E}_i$ are independent of each other. We also consider that the matrix $\mathbf{H}_i$ and the vector $\boldsymbol{\beta}_{\mathbf{x}_i}$ have elements according to the description in Section 3. Under the normality assumptions, we have $\mathbf{U}_i\sim MVN(\mathbf{0},\mathbf{D})$ and $\mathcal{E}_i\sim MVN(\mathbf{0},\mathbf{V}_i)$; consequently, $\mathbf{y}_i\sim MVN(\mathbf{H}_{i}\boldsymbol{\beta}_{\mathbf{x}_i},\mathbf{H}_{i}\mathbf{D}\mathbf{H}_{i}^{'}+\mathbf{V}_i)$ and  $\mathbf{y}_i|\mathbf{U}_i\sim MVN(\mathbf{H}_{i}\boldsymbol{\beta}_{\mathbf{x}_i}+\mathbf{H}_{i}\mathbf{U}_i,\mathbf{V}_i)$. 

Let $\ell_c(\boldsymbol{\theta}_r)$, where $\boldsymbol{\theta}_r=\{\boldsymbol{\beta}_1, \boldsymbol{\beta}_2,\ldots,\boldsymbol{\beta}_r,\boldsymbol{\xi}_{1},\boldsymbol{\xi}_{1},\ldots,\boldsymbol{\xi}_{r},\psi_1(\cdot),\psi_2(\cdot),\ldots,\psi_r(\cdot),\sigma_1^2,\sigma_2^2,\ldots, \sigma_r^2,\sigma^2\}$, be the complete data log-likelihood function induced by the model $(\ref{eq:mat_trun_model})$, using the data $\{\mathbf{y}_i,\mathbf{U}_i\}$; $i=1,2,\ldots,n$ that consists of observed $\mathbf{y}_i$ and unobserved $\mathbf{U}_i$. We denote the joint density by $f(\mathbf{y}_i,\mathbf{U}_i;\boldsymbol{\theta}_r)$ and assume independence across subjects to write 
\begin{align*}
 \ell_c(\boldsymbol{\theta}_r) =& \sum_{i=1}^{n}\log\left\{f(\mathbf{y}_i,\mathbf{U}_i;\boldsymbol{\theta}_r)\right\} \\
 =& \sum_{i=1}^{n}\log\left\{f(\mathbf{y}_i|\mathbf{U}_i;\boldsymbol{\theta}_r) f(\mathbf{U}_i;\boldsymbol{\theta}_r) \right\}\\
 \propto& \frac{pM}{2}\log (1/\sigma^2)-\frac{1}{2}\sum_{i=1}^{n}\left(\mathbf{y}_i-\mathbf{H}_{i}\boldsymbol{\beta}_{\mathbf{x}_i}-\mathbf{H}_{i}\mathbf{U}_i\right)^{'}\mathbf{V}_i^{-1}\left(\mathbf{y}_i-\mathbf{H}_{i}\boldsymbol{\beta}_{\mathbf{x}_i}-\mathbf{H}_{i}\mathbf{U}_i\right)\\
 &+\frac{n}{2} \sum_{k=1}^{r}\log(1/\sigma_k^2)-\frac{1}{2}\sum_{i=1}^n\mathbf{U}_i^{'}\mathbf{D}^{-1}\mathbf{U}_i\\
  =&\frac{pM}{2}\log (1/\sigma^2)+\frac{n}{2} \sum_{k=1}^{r}\log(1/\sigma_k^2)-\frac{1}{2}\sum_{i=1}^{n}\left\{\mathbf{y}_i^{'}\mathbf{V}_i^{-1}\mathbf{y}_i-2\mathbf{y}_i^{'}\mathbf{V}_i^{-1}\mathbf{H}_{i}\boldsymbol{\beta}_{\mathbf{x}_i}\right.\\
 &\left. +\boldsymbol{\beta}_{\mathbf{x}_i}^{'}\mathbf{H}_{i}^{'}\mathbf{V}_i^{-1}\mathbf{H}_{i}\boldsymbol{\beta}_{\mathbf{x}_i}-2\left(\mathbf{y}_i-\mathbf{H}_{i}\boldsymbol{\beta}_{\mathbf{x}_i}\right)^{'}\mathbf{V}_i^{-1}\mathbf{H}_{i}\mathbf{U}_i+\mathbf{U}_i^{'}\mathbf{H}_{i}^{'}\mathbf{V}_i^{-1}\mathbf{H}_{i}\mathbf{U}_i+\mathbf{U}_i^{'}\mathbf{D}^{-1}\mathbf{U}_i\right\}\\
=& \frac{pM}{2}\log (1/\sigma^2)+\frac{n}{2} \sum_{k=1}^{r}\log(1/\sigma_k^2)-\frac{1}{2}\sum_{i=1}^{n}\left(\mathbf{y}_i^{'}\mathbf{V}_i^{-1}\mathbf{y}_i-2\mathbf{y}_i^{'}\mathbf{V}_i^{-1}\mathbf{H}_{i}\boldsymbol{\beta}_{\mathbf{x}_i}\right.\\
&\left.+\boldsymbol{\beta}_{\mathbf{x}_i}^{'}\mathbf{H}_{i}^{'}\mathbf{V}_i^{-1}\mathbf{H}_{i}\boldsymbol{\beta}_{\mathbf{x}_i}-2\mathbf{y}_i\mathbf{V}_i^{-1}\mathbf{H}_{i}\mathbf{U}_i+2\boldsymbol{\beta}_{\mathbf{x}_i}^{'}\mathbf{H}_{i}^{'}\mathbf{V}_i^{-1}\mathbf{H}_{i}\mathbf{U}_i+\mathbf{U}_i^{'}\Gamma_{i}^{-1}\mathbf{U}_i\right)
\end{align*}
where $M=\sum_{i}^{n}m_i$ and  $\Gamma_i^{-1}=\mathbf{H}_{i}^{'}\mathbf{V}_i^{-1}\mathbf{H}_{i}+\mathbf{D}^{-1}$. In this derivation, we use expressions of $f(\mathbf{y}_i|\mathbf{U}_i;\boldsymbol{\theta}_r)$ and $f(\mathbf{U}_i;\boldsymbol{\theta}_r)$ based on $MVN(\mathbf{H}_{i}\boldsymbol{\beta}_{\mathbf{x}_i}+\mathbf{H}_{i}\mathbf{U}_i,\mathbf{V}_i)$ and  $MVN(\mathbf{0},\mathbf{D})$, respectively. 

\subsection{Conditonal mean and variance}
\label{sec:con_mean_var}

Induced by model $(\ref{eq:mat_trun_model})$, the covariance between $\mathbf{y}_i$ and $\mathbf{U}_i$ is $\mathbf{H}_i\mathbf{D}$. As $\mathbf{y}_i$ and $\mathbf{U}_i$ follow multivariate normal distributions (MVN) under the normality assumptions, $[\mathbf{U}_i,\tvec{(\mathbf{Y}_i)}]'$ also follows a MVN distribution with mean and variance, 
\begin{align*}
\left[\begin{array}{c}
     \mathbf{0}\\
     \tvec{(\mathbf{H}_i\boldsymbol{\beta}_{\mathbf{x}_i})}
\end{array}\right]\text{and}\left[\begin{array}{cc}
     D& \mathbf{D}\mathbf{H}_i^{'} \\
    \mathbf{H}_i\mathbf{D} & \mathbf{H}_i\mathbf{D}\mathbf{H}_i^{'}+\mathbf{V}_i 
\end{array}\right],
\end{align*}
respectively. Then $\mathbf{U}_i|\mathbf{y}_i$ follows a MVN with mean and covariance given as $E_{\boldsymbol{\theta}_r}(\mathbf{U}_i|\mathbf{y}_i)=\mathbf{D}\mathbf{H}_{i}^{'}\left(\mathbf{H}_{i}\mathbf{D}\mathbf{H}_{i}^{'}+\mathbf{V}_i\right)^{-1}\left(\mathbf{y}_i-\mathbf{H}_{i}\boldsymbol{\beta}_{\mathbf{x}_i}\right)$ and $V_{\boldsymbol{\theta}_r}(\mathbf{U}_i|\mathbf{y}_i)=\mathbf{D}-\mathbf{D}\mathbf{H}_{i}^{'}\left(\mathbf{H}_{i}\mathbf{D}\mathbf{H}_{i}^{'}+\mathbf{V}_i\right)^{-1}\mathbf{H}_{i}\mathbf{D}$, respectively. 

Comparing with the Woodbury matrix identity, 
\begin{align*}
(A+LCQ)^{-1}=A^{-1}-A^{-1}L(C^{-1}+QA^{-1}L)^{-1}QA^{-1},
\end{align*}
we can write 
\begin{align*}
 V_{\boldsymbol{\theta}_r}(\mathbf{U}_i|\mathbf{y}_i) = \left(\mathbf{D}^{-1}+\mathbf{H}_i^{'}\mathbf{V}_i^{-1}\mathbf{H}_i\right)^{-1}=\Gamma_i
\end{align*}
by taking $A^{-1}=\mathbf{D}$, $L=\mathbf{H}_i^{'}$, $C^{-1}=\mathbf{V}_i$ and $Q=\mathbf{H}_i$. 
To simplify $E_{\boldsymbol{\theta}_r}(\mathbf{U}_i|\mathbf{y}_i)$, we write
\begin{align*}
\left(\mathbf{H}_{i}\mathbf{D}\mathbf{H}_{i}^{'}+\mathbf{V}_i\right)^{-1}=\mathbf{V}_i^{-1}-\mathbf{V}_i^{-1}\mathbf{H}_{i}\left(\mathbf{I}+\mathbf{D}\mathbf{H}_{i}^{'}\mathbf{V}_i^{-1}\mathbf{H}_{i}\right)^{-1}\mathbf{D}\mathbf{H}_{i}^T\mathbf{V}_i^{-1}.
\end{align*}
by considering $L=\mathbf{H}_{i}$, $C=\mathbf{I}_k$,  $Q=\mathbf{D}\mathbf{H}_{i}^{'}$ and $A=\mathbf{V}_i$, where $\mathbf{I}_k$ is a $k\times k$ identity matrix. 
Finally, we obtain the simplified expression as
\begin{align*}
 E_{\boldsymbol{\theta}_r}(\mathbf{U}_i|\mathbf{y}_i)
 &=\mathbf{D}\mathbf{H}_{i}^{'}\left\{\mathbf{V}_i^{-1}-\mathbf{V}_i^{-1}\mathbf{H}_{i}\left(\mathbf{I}+\mathbf{D}\mathbf{H}_{i}^{'}\mathbf{V}_i^{-1}\mathbf{H}_{i}\right)^{-1}\mathbf{D}\mathbf{H}_{i}^{'}\mathbf{V}_i^{-1}\right\}\left(\mathbf{y}_i-\mathbf{H}_{i}\boldsymbol{\beta}_{\mathbf{x}_i}\right)\\
 &=\left\{\mathbf{D}\mathbf{H}_{i}^{'}\mathbf{V}_i^{-1}-\mathbf{D}\mathbf{H}_{i}^{'}\mathbf{V}_i^{-1}\mathbf{H}_{i}\left(\mathbf{I}+\mathbf{D}\mathbf{H}_{i}^{'}\mathbf{V}_i^{-1}\mathbf{H}_{i}\right)^{-1}\mathbf{D}\mathbf{H}_{i}^{'}\mathbf{V}_i^{-1}\right\}\left(\mathbf{y}_i-\mathbf{H}_{i}\boldsymbol{\beta}_{\mathbf{x}_i}\right)\\
 &=\left\{\mathbf{D}-\mathbf{D}\mathbf{H}_{i}^{'}\mathbf{V}_i^{-1}\mathbf{H}_{i}\left(\mathbf{I}+\mathbf{D}\mathbf{H}_{i}^{'}\mathbf{V}_i^{-1}\mathbf{H}_{i}\right)^{-1}\mathbf{D}\right\}\mathbf{H}_{i}^{'}\mathbf{V}_i^{-1}\left(\mathbf{y}_i-\mathbf{H}_{i}\boldsymbol{\beta}_{\mathbf{x}_i}\right)\\
 &=\left(\mathbf{D}^{-1}+\mathbf{H}_{i}^{'}\mathbf{V}_i^{-1}\mathbf{H}_{i}\right)^{-1}\mathbf{H}_{i}^{'}\mathbf{V}_i^{-1}\left(\mathbf{y}_i-\mathbf{H}_{i}\boldsymbol{\beta}_{\mathbf{x}_i}\right)
\end{align*}
by taking $\mathbf{L}=\mathbf{H}_{i}^{'}\mathbf{V}_i^{-1}\mathbf{H}_{i}, \mathbf{Q}=\mathbf{I},\text{ and } \mathbf{A}=\mathbf{D}^{-1}$ for the last equality. 

\subsection{M-step formulas}
\label{ssec:mstep_formula}
At $(s+1)$th iteration, the M-step deals with obtaining an updated estimate by 
\begin{align}
\label{seq:mstep_max}
\widehat{\boldsymbol{\theta}}^{(s+1)}_r=\underset{\boldsymbol{\theta}_r}{\arg\max}\text{ }Q(\boldsymbol{\theta}_r;\mathbf{y}_i,\boldsymbol{\theta}_r^{(s)}),    
\end{align}
where
\begin{align*}
Q(\boldsymbol{\theta}_r;\mathbf{y}_i,\boldsymbol{\theta}_r^{(s)})&=\frac{pM}{2}\log (1/\sigma^2)+\frac{n}{2}\sum_{k=1}^{r}\log(1/\sigma_k^2) -\frac{1}{2}\sum_{i=1}^{n}\left[
    \left|\left|\mathbf{y}_i-\mathbf{H}_{i}\boldsymbol{\beta}_{\mathbf{x}_i}-\mathbf{H}_{i}\widetilde{\mathbf{U}}_i^{(s)}\right|\right|_{\mathbf{V}_i}^{2}\right.\\
&\left.+\text{tr}\left\{\left(\mathbf{H}_{i}^{'}\mathbf{V}_i^{-1}\mathbf{H}_{i}\right)\Gamma_i^{(s)}\right\}+\text{tr}\left(\mathbf{D}^{-1}\Gamma_i^{(s)}\right)+\left|\left|\widetilde{\mathbf{U}}_i^{(s)}\right|\right|_{\mathbf{D}}^{2}\right]+\sum_{k=1}^{r}\eta_k||\psi_k||_{\mathcal{H}}.
\end{align*}
In the implementation, we construct two matrices $\mathbf{R}_{i,k}^{(s)}$ and $\widetilde{\mathbf{R}}_{i,k}^{(s)}$, each of dimension $p\times m_i$, with elements 
\begin{align*}
 R_{ijk}^{b,(s)}&=Y_{ij}^{b}-\sum_{l\ne k=1}^{r}\left\{\widetilde{U}_{ik}^{(s)}+\mathbf{x}_i^{'}\widehat{\boldsymbol{\beta}}_{l}^{(s)}\right\}\widehat{\xi}_{bl}^{(s)}\widehat{\psi}_{l}^{(s)}(t_{ij}),\text{ and }\\
 \widetilde{R}_{ijk}^{b,(s)}&= \frac{R_{ijk}^{b,(s)}\left\{\widetilde{U}_{ik}^{(s)}+\mathbf{x}_i^{'}\widehat{\boldsymbol{\beta}}_k^{(s+1)}\right\}-\sum_{l\ne k=1}^{r}\widehat{\xi}_{bl'}^{(s)}\widehat{\psi}_{l}^{(s)}(t_{ij})\Gamma_{i,kl}^{(s)}}{\sqrt{\left\{\widetilde{U}_{ik}^{(s)}+\mathbf{x}_i^{'}\widehat{\boldsymbol{\beta}}_k^{(s+1)}\right\}^2+\Gamma_{i,kk}^{(s)}}}
\end{align*}
at cell $(b,j)$, respectively. The maximization  $(\ref{seq:mstep_max})$ with respect to the parameter of interest, $\boldsymbol{\beta}_k$, $\boldsymbol{\xi}_k$ and $\psi_k(\cdot)$, reduces to separate least square problems as
\begin{align}
\label{seq:mstep_opt}
\begin{array}{rl}
    (a)&\underset{\boldsymbol{\beta}_k}{\min}\left|\left|\tvec{(\mathbf{R}_{i,k}^{(s)})}-\tvec{(\widehat{\boldsymbol{\xi}}_k^{(s)}\circ\widehat\Psi_{ik}^{(s)})}(\widetilde U_{ik}^{(s)}+\mathbf{x}_i^{'}\boldsymbol{\beta}_k)\right|\right|_2^2\\
    (b)&\underset{\boldsymbol{\xi}_k}{\min}\left|\left|\tvec{(\widetilde{\mathbf{R}}_{i,k}^{(s)})}-\tvec{(\boldsymbol{\xi}_k\circ\widehat\Psi_{ik}^{(s)})}\sqrt{(\widetilde U_{ik}^{(s)}+\mathbf{x}_i^{'}\widehat{\boldsymbol{\beta}}_k^{(s+1)})^2+\Gamma_{i,kk}^{(s)}}\right|\right|_2^2\\    
    (c)&\underset{\psi_k\in L^2(\mathcal{T})}{\min}\left|\left|\tvec{(\widetilde{\mathbf{R}}_{i,k}^{(s)})}-\tvec{(\widehat{\boldsymbol{\xi}}_k^{(s+1)}\circ\Psi_{ik})}\sqrt{(\widetilde U_{ik}^{(s)}+\mathbf{x}_i^{'}\widehat{\boldsymbol{\beta}}_k^{(s+1)})^2+\Gamma_{i,kk}^{(s)}}\right|\right|_2^2+\eta_k\left|\left|\phi_k\right|\right|_{\mathcal{H}}\\    
\end{array}
\end{align}
where $||\cdot||_2$ represents the Euclidean norm of a vector, $\widehat{\Psi}_{ik}^{(s)}=[\widehat\psi_k^{(s)}(t_{i1}),\widehat\psi_k^{(s)}(t_{i2}),\ldots,\widehat\psi_k^{(s)}(t_{im_i})]^{'}$, $\Gamma_{i,kk}^{(s)}$ is the value of $\Gamma_{i}^{(s)}$ at cell $(k,k)$, and $L^2(\mathcal{T})$ is the space of real-valued squared-integrable function defined over $\mathcal{T}$. In scalar notations, we rewrite the least-squares objective functions in $(\ref{seq:mstep_opt})$ as
\begin{align}
\label{seq:mstep_opt_scl}
\begin{array}{rl}
    (a)&\underset{\boldsymbol{\beta}_k}{\min}\text{ }\sum_{i=1}^{n}\sum_{b=1}^{p}\sum_{j=1}^{m_i}\left[R_{ijk}^{b,(s)}-\widetilde{U}_{ik}^{(s)}\widehat{\xi}_{bk}^{(s)}\widehat\psi_k^{(s)}(t_{ij})-\left\{\mathbf{x}_i^{'}\boldsymbol{\beta}_{k}\right\}\widehat\xi_{bk}^{(s)}\widehat\psi_{k}^{(s)}(t_{ij})\right]^2,\\
    (b)&\underset{\xi_{bk}}{\min}\text{ }\sum_{i=1}^{n}\sum_{j=1}^{m_i}\left[\widetilde{R}_{ijk}^{b,(s)}-\sqrt{\left\{\widetilde{U}_{ik}^{(s)}+\mathbf{x}_i^{'}\widehat{\boldsymbol{\beta}}_k^{(s+1)}\right\}^2+\Gamma_{i,kk}^{(s)}}\widehat\psi_{k}^{(s)}(t_{ij})\xi_{bk}\right]^2; b=1,2,\ldots,p,\\  
    (c)&\underset{\psi\in L^2[\mathcal{T}]}{\min}\text{ }\sum_{i=1}^{n}\sum_{b=1}^{p}\sum_{j=1}^{m_i}\left[\widetilde{R}_{ijk}^{b,(s)}-\sqrt{\left\{\widetilde{U}_{ik}^{(s)}+\mathbf{x}_i^{'}\widehat{\boldsymbol{\beta}}_k^{(s+1)}\right\}^2+\Gamma_{i,kk}^{(s)}}\widehat\xi_{bk}^{(s+1)}\psi_{k}(t_{ij})\vphantom{\frac{\left\{\sum_{b=1}^{p}Y_{ijk}^{(b)}\xi_{bk}^{(s)}\right\}\left[\widehat{U}_{ik}^{(s)}+\left\{\mathbf{x}_i^{'}\boldsymbol{\beta}_k^{(s)}\right\}\right]}{\sqrt{\left\{\widehat{U}_{ik}^{(s)}+\mathbf{`x}_i^{'}\boldsymbol{\beta}_k^{(s)}\right\}^2+\text{var}\left\{U_{ik}|\tvec{(\mathbf{Y}_i)},\boldsymbol{\Theta}^{(s)}\right\}}}}\right]^2+\eta_k||\psi_k||_{\mathcal{H}}.\\    
\end{array}
\end{align}

Let $\boldsymbol{\beta}_{k}^{(s+1)}$ and $\widetilde{\xi}_{bk}^{(s+1)}; b=1,2,\ldots,p$ be the minimizers of $(p+1)$ optimization in $(a)$ and $(b)$ of $(\ref{seq:mstep_opt_scl})$, then we can write
\begin{align*}
\widehat{\boldsymbol{\beta}}_k^{(s+1)} = \left\{\mathbf{X}_k^{(s)'}\mathbf{X}_k^{(s)}\right\}^{-1}\mathbf{X}_k^{(s)'}\mathbf{Z}_k^{(s)},
\end{align*}
where $\mathbf{X}_k^{(s)}\in\mathbb{R}^{pM\times q}$ with rows specified by $\mathbf{x}_i\widehat{\xi}_{bk}^{(s)}\widehat\psi_{k}^{(s)}(t_{ij})$, $\mathbf{Z}_k^{(s)}\in\mathbb{R}^{pM\times 1}$ with elements $R_{ijk}^{b,(s)}-\widetilde{U}_{ik}^{(s)}\widehat\xi_{bk}^{(s)}\widehat\psi_k^{(s)}(t_{ij})$, and  \begin{align*}
\widetilde{\xi}_{bk}^{(s+1)} &= \frac{\sum_{i=1}^{n}\sum_{j=1}^{m_i}\widetilde{R}_{ijk}^{b,(s)}\sqrt{\left\{\widetilde{U}_{ik}^{(s)}+\mathbf{x}_i^{'}\widehat{\boldsymbol{\beta}}_k^{(s+1)}\right\}^2+\Gamma_{i,kk}^{(s)}}\widehat\psi_{k}^{(s)}(t_{ij})}{\sum_{i=1}^{n}\sum_{j=1}^{m_i}\left[\sqrt{\left\{\widetilde{U}_{ik}^{(s)}+\mathbf{x}_i^{'}\widehat{\boldsymbol{\beta}}_k^{(s+1)}\right\}^2+\Gamma_{i,kk}^{(s)}}\widehat\psi_{k}^{(s)}(t_{ij})\right]^2}; b=1,2,\ldots,p.
\end{align*}
We define the minimizer $\widehat{\boldsymbol{\xi}}_k^{(s+1)}=\widetilde{\boldsymbol{\xi}}_k^{(s+1)}/||\widetilde{\boldsymbol{\xi}}_k^{(s+1)}||_2$, where $\widetilde{\boldsymbol{\xi}}_k^{(s+1)}=(\widetilde{\xi}_{1k}^{(s+1)}, \widetilde{\xi}_{2k}^{(s+1)}, \ldots, \widetilde{\xi}_{pk}^{(s+1)})^{'}$. For updating the $(r+1)$ variance parameters, we also have exact expressions, which are
\begin{align*}
\widehat\sigma_k^{2,(s+1)}&=\frac{1}{n}\sum_{i=1}^{n}\left(\widetilde{U}_{ik}^{(s)}\widetilde{U}_{ik}^{(s)}+\Gamma_{i,kk}^{(s)}\right); k=1,2,\ldots,r, \text{and}\\
\widehat\sigma^{2,(s+1)}&=\frac{1}{pM}\sum_{i=1}^{n}\sum_{b=1}^{p}\sum_{j=1}^{m_i}\sum_{k=1}^{r}\sum_{l=1}^{r}\left\{\widehat\xi_{bk}^{(s+1)}\widehat\xi_{bl}^{(s+1)}\widehat\psi_k^{(s+1)}(t_{ij})\widehat\psi_l^{(s+1)}(t_{ij})\Gamma_{i,kl}^{(s)}\right\}\\
&+\frac{1}{pM}\sum_{i=1}^{n}\sum_{b=1}^{p}\sum_{j=1}^{m_i}\left\{Y_{ijk}^{b}-\sum_{k=1}^{r}\left(\mathbf{x}_i^{'}\widehat{\boldsymbol{\beta}}_k^{(s+1)}+\widetilde{U}_{ik}^{(s)}\right)\widehat\xi_{bk}^{(s+1)}\widehat\psi_{k}^{(s+1)}(t_{ij})\right\}^2.
\end{align*}

\end{document}